\documentclass[prd,preprintnumbers,twocolumn,amsmath,nofootinbib,amssymb]{revtex4}
\usepackage{graphicx,color,dcolumn,booktabs,bm}
\usepackage{longtable,lscape}
\usepackage{txfonts}
\usepackage{overpic}
\usepackage{amssymb}
\usepackage{epstopdf}
\usepackage{indentfirst}
\usepackage{feynmf}   
\usepackage{slashed}  
\usepackage{cases}
\usepackage{color}
\usepackage{float}
\usepackage{multirow}
\usepackage{ulem}
\usepackage{graphicx,color,dcolumn,booktabs,bm}
\usepackage{epsfig,dsfont,amssymb,amsmath,amsfonts,amsbsy,mathrsfs}

\graphicspath{{Figures/}} %

\usepackage{hyperref}
\hypersetup{colorlinks,citecolor=blue,anchorcolor=red,menucolor=red, linkcolor=red,filecolor=red,runcolor=red,urlcolor=blue,frenchlinks=red}

\makeatletter
\@addtoreset{equation}{section}
\makeatother

\allowdisplaybreaks

\begin{document}

\title{New type of doubly charmed molecular pentaquarks containing most strange quarks: Mass spectra, radiative decays, and magnetic moments}

\author{Fu-Lai Wang$^{1,2,3,5}$}
\email{wangfl2016@lzu.edu.cn}
\author{Xiang Liu$^{1,2,3,4,5}$}
\email{xiangliu@lzu.edu.cn}
\affiliation{$^1$School of Physical Science and Technology, Lanzhou University, Lanzhou 730000, China\\
$^2$Lanzhou Center for Theoretical Physics, Key Laboratory of Theoretical Physics of Gansu Province, Lanzhou University, Lanzhou 730000, China\\
$^3$Key Laboratory of Quantum Theory and Applications of MoE, Lanzhou University,
Lanzhou 730000, China\\
$^4$MoE Frontiers Science Center for Rare Isotopes, Lanzhou University, Lanzhou 730000, China\\
$^5$Research Center for Hadron and CSR Physics, Lanzhou University and Institute of Modern Physics of CAS, Lanzhou 730000, China}

\begin{abstract}
In this work, we first predict the mass spectra of the $\Omega_{c}^{(*)}{D}_s^{(*)}$-type doubly charmed molecular pentaquark candidates, where the one-boson-exchange model is adopted by considering both the $S$-$D$ wave mixing effect and the coupled channel effect. Our findings indicate that the $\Omega_{c}{D}_s^*$ state with $J^P={1}/{2}^{-}$, the $\Omega_{c}^*{D}_s^*$ state with $J^P={1}/{2}^{-}$, and the $\Omega_{c}^*{D}_s^*$ state with $J^P={3}/{2}^{-}$ can be considered as the most promising doubly charmed molecular pentaquark candidates, and the $\Omega_{c}{D}_s$ state with $J^P={1}/{2}^{-}$, the $\Omega_{c}^*{D}_s$ state with $J^P={3}/{2}^{-}$, and the $\Omega_{c}{D}_s^*$ state with $J^P={3}/{2}^{-}$ are the possible doubly charmed molecular pentaquark candidates. Furthermore, we further explore the radiative decays and the magnetic moments of the most promising doubly charmed molecular pentaquark candidates in the constituent quark model. As a crucial aspect of spectroscopy, the information of the radiative decays and the magnetic moments can provide the valuable clues to reflect their inner structures. With the accumulation of higher statistical data at the Large Hadron Collider, we propose that the LHCb Collaboration should focus on the problem of searching for these predicted doubly charmed molecular pentaquark candidates containing most strange quarks in the coming years.
\end{abstract}

\maketitle

\section{Introduction}\label{sec1}

As an important and intriguing research topic, the study of the exotic hadronic states is one of the effective approaches to further deepen our understanding of the non-perturbative behavior of the strong interaction. As is well known, the investigation of the exotic hadronic states has been a long history in hadron physics, dating back to the inception of the quark model in 1964 by M. Gell-Mann and G. Zweig \cite{GellMann:1964nj,Zweig:1981pd}. In particular, since the discovery of the charmonium-like state $X(3872)$ by the Belle Collaboration in 2003 \cite{Choi:2003ue}, the high-energy physics experiments have reported more and more new hadronic states, which prompts extensive discussions on the existence of the hadronic molecular states, the compact multiquark states, the glueballs, the hybrids, and so on \cite{Liu:2013waa,Hosaka:2016pey,Chen:2016qju,Richard:2016eis,Lebed:2016hpi,Brambilla:2019esw,Liu:2019zoy,Chen:2022asf,Olsen:2017bmm,Guo:2017jvc,Meng:2022ozq}. Among them, the hadronic molecular states have frequently been employed to discuss the observed new hadronic states in the last two decades \cite{Liu:2013waa,Hosaka:2016pey,Chen:2016qju,Richard:2016eis,Lebed:2016hpi,Brambilla:2019esw,Liu:2019zoy,Chen:2022asf,Olsen:2017bmm,Guo:2017jvc,Meng:2022ozq}, and this is due to the fact that the majority of the newly observed hadronic states are located near the thresholds of the corresponding two hadronic states.

An important example is the observation of the $P_{\psi}^{N}(4312)$, $P_{\psi}^{N}(4440)$, and $P_{\psi}^{N}(4457)$ states in the $J/\psi p$ invariant mass spectrum by the LHCb Collaboration in 2019 \cite{Aaij:2019vzc}, and their masses are below and close to the corresponding thresholds of the $S$-wave charmed baryon $\Sigma_c$ and the $S$-wave anti-charmed meson $\bar D^{(*)}$, which provides the compelling experimental evidence for the existence of the $\Sigma_c \bar D^{(*)}$-type hidden-charm molecular pentaquark states in nature \cite{Li:2014gra,Karliner:2015ina,Wu:2010jy,Wang:2011rga,Yang:2011wz,Wu:2012md,Chen:2015loa}. In recent years, LHCb also found the evidence for the existence of the $P_{\psi s}^{\Lambda}(4459)$ state \cite{LHCb:2020jpq} and reported the discovery of the $P_{\psi s}^{\Lambda}(4338)$ state \cite{LHCb:2022ogu} in the $J/\psi\Lambda$ invariant mass spectrum, which are associated with the $\Xi_c \bar D^{(*)}$-type hidden-charm molecular pentaquark states with strangeness \cite{Chen:2016ryt,Wu:2010vk,Hofmann:2005sw,Anisovich:2015zqa,Feijoo:2015kts,Lu:2016roh,Xiao:2019gjd,Chen:2015sxa,Wang:2019nvm,Weng:2019ynv}, respectively. Moreover, LHCb announced the observation of the doubly charmed tetraquark state $T_{cc}(3875)^+$ in the $D^0D^0\pi^+$ invariant mass spectrum in 2021 \cite{LHCb:2021vvq}, and several theoretical groups have suggested the observed $T_{cc}(3875)^+$ state assignments to the $D D^{*}$-type doubly charmed molecular tetraquark state \cite{Manohar:1992nd,Ericson:1993wy,Tornqvist:1993ng,Janc:2004qn,Ding:2009vj,Molina:2010tx,Ding:2020dio,Li:2012ss,Xu:2017tsr,Liu:2019stu,Ohkoda:2012hv,Tang:2019nwv}.

Building on the observations of the hidden-charm molecular pentaquark candidates $P_{\psi}^{N}/P_{\psi s}^{\Lambda}$ \cite{Aaij:2019vzc,LHCb:2020jpq,LHCb:2022ogu} and the doubly charmed molecular tetraquark candidate $T_{cc}$ \cite{LHCb:2021vvq} in recent years, it is natural to anticipate the existence of a family of the doubly charmed molecular pentaquarks in the hadron spectroscopy, which contains different types of doubly charmed molecular pentaquark states. In Refs. \cite{Chen:2021kad,Yalikun:2023waw}, the authors have already predicted the existence of the doubly charmed molecular pentaquark candidates with the $\Sigma_c^{(*)} D^{(*)}$ and $\Xi_c^{(\prime,*)} D_s^{(*)}$ configurations based on the one-boson-exchange (OBE) model by introducing the $S$-$D$ wave mixing effect and the coupled channel effect. Given the above information, it is natural to focus on the mass spectra and the properties of the doubly charmed molecular pentaquark candidates with the $\Omega_{c}^{(*)}{D}_s^{(*)}$ configuration to further complete the family of the doubly charmed molecular pentaquarks, which are new type of doubly charmed molecular pentaquarks containing most strange quarks.

In the present work, we first predict the mass spectra of the $\Omega_{c}^{(*)}{D}_s^{(*)}$-type doubly charmed molecular pentaquark candidates. The OBE model is utilized to obtain the effective interactions of the $\Omega_{c}^{(*)}{D}_s^{(*)}$ systems, which is a reliable tool to estimate the hadron-hadron interactions \cite{Chen:2016qju, Liu:2019zoy}. Based on the obtained OBE effective interactions, the bound state properties of the $\Omega_{c}^{(*)}{D}_s^{(*)}$ systems can be discussed by solving the coupled channel Schr\"{o}dinger equation, and the realistic calculations consider both the $S$-$D$ wave mixing effect and the coupled channel effect, which can provide the important information for predicting the mass spectra of the $\Omega_{c}^{(*)}{D}_s^{(*)}$-type doubly charmed molecular pentaquark candidates. Meanwhile, our calculations yield the  corresponding spatial wave functions of the $\Omega_{c}^{(*)}{D}_s^{(*)}$ molecules. Furthermore, we further discuss the radiative decays and the magnetic moments of the obtained most promising doubly charmed molecular pentaquark candidates based on their mass spectra and spatial wave functions, where we utilise the constituent quark model, which has been widely used for studying the magnetic moments and the radiative decays of the hadrons in recent decades \cite{Liu:2003ab,Huang:2004tn,Zhu:2004xa,Haghpayma:2006hu,Wang:2016dzu,Deng:2021gnb,Gao:2021hmv,Zhou:2022gra,Wang:2022tib,Li:2021ryu,Schlumpf:1992vq,Schlumpf:1993rm,Cheng:1997kr,Ha:1998gf,Ramalho:2009gk,Girdhar:2015gsa,Menapara:2022ksj,Mutuk:2021epz,Menapara:2021vug,Menapara:2021dzi,Gandhi:2018lez,Dahiya:2018ahb,Kaur:2016kan,Thakkar:2016sog,Shah:2016vmd,Dhir:2013nka,Sharma:2012jqz,Majethiya:2011ry,Sharma:2010vv,Dhir:2009ax,Simonis:2018rld,Ghalenovi:2014swa,Kumar:2005ei,Rahmani:2020pol,Hazra:2021lpa,Gandhi:2019bju,Majethiya:2009vx,Shah:2016nxi,Shah:2018bnr,Ghalenovi:2018fxh,Wang:2022nqs,Mohan:2022sxm,An:2022qpt,Kakadiya:2022pin,Wu:2022gie,Wang:2023bek}. This study seeks to provide more abundant suggestions to encourage our experimental colleagues to explore the $\Omega_{c}^{(*)}{D}_s^{(*)}$-type doubly charmed molecular pentaquark candidates in the forthcoming experiments, which is the crucial advancement towards in constructing the family of the doubly charmed molecular pentaquarks. Although no doubly charmed molecular pentaquark candidates have been reported to date, it is likely that the doubly charmed molecular pentaquark candidates will be detected by LHCb with the accumulation of higher statistical data at the Large Hadron Collider (LHC) \cite{Bediaga:2018lhg}.

The rest of this paper is structured as follows. In Sec. \ref{sec2}, we give the derivation of the OBE effective potentials for the $\Omega_{c}^{(*)}{D}_s^{(*)}$ systems and predict the mass spectra of the $\Omega_{c}^{(*)}{D}_s^{(*)}$-type doubly charmed molecular pentaquark candidates. Utilizing the obtained mass spectra and spatial wave functions of the $\Omega_{c}^{(*)}{D}_s^{(*)}$-type doubly charmed molecular pentaquark candidates, we further explore the radiative decays and the magnetic moments of the most promising doubly charmed molecular pentaquark candidates within the constituent quark model in Sec. \ref{sec3}. Finally, the summary is given in Sec. \ref{sec4}.

\section{The prediction of mass spectra of the $\Omega_{c}^{(*)}{D}_s^{(*)}$ molecules}\label{sec2}

This section concentrates on the prediction of the mass spectra for the $\Omega_{c}^{(*)}{D}_s^{(*)}$-type doubly charmed molecular pentaquark candidates, which can offer the essential information for the experimental search for them. Meanwhile, our calculations yield the corresponding spatial wave functions of the $\Omega_{c}^{(*)}{D}_s^{(*)}$ molecules, which are the significant inputs for investigating their properties. In the following, we first present the derivation of the OBE effective potentials for the $\Omega_{c}^{(*)}{D}_s^{(*)}$ systems. And then, we analyse the bound state properties of the $\Omega_{c}^{(*)} D_s^{(*)}$ systems by solving the coupled channel Schr\"{o}dinger equation, which provides us the critical information for predicting the mass spectra of the $\Omega_{c}^{(*)}{D}_s^{(*)}$-type doubly charmed molecular pentaquark candidates.

\subsection{The derivation of the interactions of the $\Omega_{c}^{(*)}{D}_s^{(*)}$ systems within the OBE model}

As is well known, the interactions between the constituent hadrons play a crucial role in studying the mass spectra of the hadronic molecular states. This study employs the OBE model to derive the effective potentials in the coordinate space of the $\Omega_{c}^{(*)}{D}_s^{(*)}$ systems \cite{Chen:2016qju, Liu:2019zoy}, which has frequently been utilized in the exploration of the observed $P_{\psi}^{N}$ \cite{Aaij:2019vzc}, $P_{\psi s}^{\Lambda}$ \cite{LHCb:2020jpq,LHCb:2022ogu}, and $T_{cc}$ \cite{LHCb:2021vvq} within the hadronic molecular picture and predict the hadronic molecular candidates \cite{Chen:2016qju, Liu:2019zoy}.

\subsubsection{Introducing the flavor and spin-orbital wave functions of the $\Omega_{c}^{(*)}{D}_s^{(*)}$ systems}

For the $\Omega_{c}^{(*)}{D}_s^{(*)}$ systems, the flavor wave functions $|I,I_3\rangle$ have the form of $|0,0\rangle=|\Omega_{c}^{(*)0}{D}_s^{(*)+}\rangle$, where $I$ and $I_3$ correspond to the isospin and the isospin third component of the studied system.

In addition, the spin-orbital wave functions $|{}^{2S+1}L_{J}\rangle$ for the $\Omega_{c}^{(*)}{D}_s^{(*)}$ systems are
\begin{eqnarray}
\left|{}^{2S+1}L_{J}\right\rangle^{\Omega_{c}{D}_{s}}&=&\sum_{m,m_L}C^{J,M}_{\frac{1}{2}m,Lm_L}\chi_{\frac{1}{2}m}Y_{L m_L},\nonumber\\
\left|{}^{2S+1}L_{J}\right\rangle^{\Omega_{c}^*{D}_{s}}&=&\sum_{m,m_L}C^{J,M}_{\frac{3}{2}m,Lm_L}\Phi_{\frac{3}{2}m}Y_{L m_L},\nonumber\\
\left|{}^{2S+1}L_{J}\right\rangle^{\Omega_{c}{D}_{s}^*}&=&\sum_{m,m',m_S,m_L}C^{S,m_S}_{\frac{1}{2}m,1m'}C^{J,M}_{Sm_S,Lm_L}\chi_{\frac{1}{2}m}\epsilon_{m'}^{\mu}Y_{L m_L},\nonumber\\
\left|{}^{2S+1}L_{J}\right\rangle^{\Omega_{c}^{*}{D}_{s}^*}&=&\sum_{m,m',m_S,m_L}C^{S,m_S}_{\frac{3}{2}m,1m'}C^{J,M}_{Sm_S,Lm_L}\Phi_{\frac{3}{2}m}^{\mu}\epsilon_{m'}^{\nu}Y_{L m_L}.\nonumber\\
\end{eqnarray}
Here, $S$, $L$, and $J$ in the spin-orbital wave functions $|{}^{2S+1}L_{J}\rangle$ represent the spin, the orbital angular momentum, and the total angular momentum for the $\Omega_{c}^{(*)}{D}_s^{(*)}$ systems, respectively. Under the static limit, we can explicitly express the polarization vector $\epsilon_{m}^{\mu}\,(m=0,\,\pm1)$ as $\epsilon_{\pm}^{\mu}= \left(0,\,\pm1,\,i,\,0\right)/\sqrt{2}$ and $\epsilon_{0}^{\mu}= \left(0,0,0,-1\right)$. $\chi_{\frac{1}{2}m}$ refers to the spin wave function with the spin $S={1}/{2}$, and the polarization tensor $\Phi_{\frac{3}{2}m}^{\mu}$ with the spin $S={3}/{2}$ can be represented explicitly as $\Phi_{\frac{3}{2}m}^{\mu}=\sum_{m_1,m_2}C^{\frac{3}{2},m}_{\frac{1}{2}m_1,1m_2}\chi_{\frac{1}{2}m_1}\epsilon_{m_2}^{\mu}$, where the Clebsch-Gordan coefficient is denoted by the constant $C^{e,f}_{ab,cd}$. In addition, $Y_{L m_L}$ is the spherical harmonic function.

In our concrete calculations, the contribution of the $S$-$D$ wave mixing effect and the coupled channel effect is considered when discussing the $\Omega_{c}^{(*)}{D}_s^{(*)}$ interactions, where the relevant $S$-wave and $D$-wave channels $|{}^{2S+1}L_{J}\rangle$ of the $\Omega_{c}^{(*)}{D}_s^{(*)}$ systems include
\begin{eqnarray}
J^P=\frac{1}{2}^-:&&\Omega_{c} D_s|{}^2\mathbb{S}_{\frac{1}{2}}\rangle,\quad\Omega_{c} D_s^*|{}^2\mathbb{S}_{\frac{1}{2}}/{}^4\mathbb{D}_{\frac{1}{2}}\rangle,\nonumber\\
&&\Omega_{c}^* D_s^*|{}^2\mathbb{S}_{\frac{1}{2}}/{}^4\mathbb{D}_{\frac{1}{2}}/{}^6\mathbb{D}_{\frac{1}{2}}\rangle,\nonumber\\
J^P=\frac{3}{2}^-:&&\Omega_{c}^* D_s|{}^4\mathbb{S}_{\frac{3}{2}}/{}^4\mathbb{D}_{\frac{3}{2}}\rangle,\quad\Omega_{c} D_s^*|{}^4\mathbb{S}_{\frac{3}{2}}/{}^2\mathbb{D}_{\frac{3}{2}}/{}^4\mathbb{D}_{\frac{3}{2}}\rangle,\nonumber\\
&&\Omega_{c}^* D_s^*|{}^4\mathbb{S}_{\frac{3}{2}}/{}^2\mathbb{D}_{\frac{3}{2}}/{}^4\mathbb{D}_{\frac{3}{2}}/{}^6\mathbb{D}_{\frac{3}{2}}\rangle,\nonumber\\
J^P=\frac{5}{2}^-:&&\Omega_{c}^* D_s^*|{}^6\mathbb{S}_{\frac{5}{2}}/{}^2\mathbb{D}_{\frac{5}{2}}/{}^4\mathbb{D}_{\frac{5}{2}}/{}^6\mathbb{D}_{\frac{5}{2}}\rangle.
\end{eqnarray}
Here, we take $|^{2S+1}L_J\rangle$ to represent the spin $S$, the orbital angular momentum $L$, and the total angular momentum $J$ of the corresponding mixing channels, respectively.

\subsubsection{Deducing the interactions of the $\Omega_{c}^{(*)}{D}_s^{(*)}$ systems within the OBE model}

For the $\Omega_{c}^{(*)}{D}_s^{(*)}$ systems, the $f_0(980)$, $\eta$, and $\phi$ exchange interactions are taken into account within the OBE model. To simplify, the meson $f_0(980)$ will henceforth be referred to as $f_0$.

To quantitatively derive the $\Omega_{c}^{(*)}{D}_s^{(*)}$ interactions at the hadronic level, we first need to compute the scattering amplitude $\mathcal{M}^{\Omega_{c}^{(*)}{D}_s^{(*)}\to \Omega_{c}^{(*)}{D}_s^{(*)}}(\bm{q})$ for the $\Omega_{c}^{(*)}{D}_s^{(*)}\to \Omega_{c}^{(*)}{D}_s^{(*)}$ process by employing the effective Lagrangian approach \cite{LHCb:2021vvq}. In detail, the scattering amplitude $\mathcal{M}^{\Omega_{c}^{(*)}{D}_s^{(*)}\to \Omega_{c}^{(*)}{D}_s^{(*)}}(\bm{q})$ of the $\Omega_{c}^{(*)}{D}_s^{(*)}\to \Omega_{c}^{(*)}{D}_s^{(*)}$ process can be calculated by the relation  $i\mathcal{M}^{\Omega_{c}^{(*)}{D}_s^{(*)}\to \Omega_{c}^{(*)}{D}_s^{(*)}}(\bm{q})=\sum_{\mathcal{E}=f_0,\,\mathbb{P},\,\mathbb{V}}i\Gamma^{\Omega_{c}^{(*)}\Omega_{c}^{(*)}\mathcal{E}}_{(\mu)} P_{\mathcal{E}}^{(\mu\nu)} i\Gamma^{{D}_s^{(*)}{D}_s^{(*)}\mathcal{E}}_{(\nu)}$
when considering the exchange of the light mesons $\mathcal{E}$ \cite{Wang:2021ajy,Wang:2023ftp}. Here, the propagator of the light meson $\mathcal{E}$ is denoted by $P_{\mathcal{E}}^{(\mu\nu)}$, while the associated interaction vertices for the $\Omega_{c}^{(*)}{D}_s^{(*)}\to \Omega_{c}^{(*)}{D}_s^{(*)}$ scattering process by exchanging the light meson $\mathcal{E}$ are given as $\Gamma^{\Omega_{c}^{(*)}\Omega_{c}^{(*)}\mathcal{E}}_{(\mu)}$ and $\Gamma^{{D}_s^{(*)}{D}_s^{(*)}\mathcal{E}}_{(\nu)}$. In the following, we construct the relevant effective Lagrangians to describe the effective interactions of the $\Omega_{c}^{(*)} D_s^{(*)}$ systems.

Regarding the constituent hadrons of the $\Omega_{c}^{(*)} D_s^{(*)}$ systems, $\Omega_c$ with $J^P=1/2^+$ and $\Omega_c^*$ with $J^P=3/2^+$ are part of the $S$-wave single-charmed baryons in the $6_F$ flavor representation, and previous theoretical studies commonly take the notation $\mathcal{B}_6^{(*)}$ to refer to the matrix of the $S$-wave single-charmed baryons in the $6_F$ flavor representation, which can be expressed as \cite{Wise:1992hn,Casalbuoni:1992gi,Casalbuoni:1996pg,Yan:1992gz,Bando:1987br,Harada:2003jx,Chen:2017xat}
\begin{eqnarray}
\mathcal{B}_6^{(*)} = \left(\begin{array}{ccc}
         \Sigma_c^{{(*)}++}                  &\frac{1}{\sqrt{2}}\Sigma_c^{{(*)}+}     &\frac{1}{\sqrt{2}}\Xi_c^{(\prime,*)+}\\
         \frac{1}{\sqrt{2}}\Sigma_c^{{(*)}+}      &\Sigma_c^{{(*)}0}    &\frac{1}{\sqrt{2}}\Xi_c^{(\prime,*)0}\\
         \frac{1}{\sqrt{2}}\Xi_c^{(\prime,*)+}    &\frac{1}{\sqrt{2}}\Xi_c^{(\prime,*)0}     &\Omega_c^{(*)0}
\end{array}\right).
\end{eqnarray}
In addition, $D_s$ with $J^P=0^-$ and $D_s^*$ with $J^P=1^-$ are the $S$-wave charmed-strange mesons \cite{ParticleDataGroup:2022pth}. For the exchanged light mesons for the $\Omega_{c}^{(*)}D_s^{(*)}$ systems, we consider the $f_0$, $\eta$, and $\phi$ mesons, which are the light scalar, pseudoscalar, and vector mesons \cite{ParticleDataGroup:2022pth}, respectively.

The effective Lagrangians describing the interactions between the heavy hadrons $\mathcal{B}_6^{(*)}/{D}_s^{(*)}$ and the light scalar, pseudoscalar, and vector mesons are constructed by incorporating the requirements of the heavy quark symmetry, the chiral symmetry, and the hidden local symmetry \cite{Wise:1992hn,Casalbuoni:1992gi,Casalbuoni:1996pg,Yan:1992gz,Bando:1987br,Harada:2003jx,Chen:2017xat,Ding:2008gr}, i.e.,
\begin{eqnarray}
\mathcal{L}_{\mathcal{B}^{(*)}_6\mathcal{B}^{(*)}_6\mathcal{E}} &=&  l_S\langle\bar{\mathcal{S}}_{\mu}f_0\mathcal{S}^{\mu}\rangle
         -\frac{3}{2}g_1\varepsilon^{\mu\nu\lambda\kappa}v_{\kappa}\langle\bar{\mathcal{S}}_{\mu}{\mathcal A}_{\nu}\mathcal{S}_{\lambda}\rangle\nonumber\\
  &&+i\beta_{S}\langle\bar{\mathcal{S}}_{\mu}v_{\alpha}\left(\mathcal{V}^{\alpha}-\rho^{\alpha}\right) \mathcal{S}^{\mu}\rangle\nonumber\\
  &&+\lambda_S\langle\bar{\mathcal{S}}_{\mu}F^{\mu\nu}(\rho)\mathcal{S}_{\nu}\rangle,\\
\mathcal{L}_{{D}_s^{(*)}{D}_s^{(*)}\mathcal{E}}&=&g_{S}\left\langle H^{(Q)}_af_0\bar{H}^{(Q)}_a\right\rangle+ig\left\langle H^{(Q)}_b{\mathcal A}\!\!\!\slash_{ba}\gamma_5\bar{H}^{\,({Q})}_a\right\rangle\nonumber\\
  &&+i\beta\left\langle H^{(Q)}_b v^{\mu}({\mathcal V}_{\mu}-\rho_{\mu})_{ba}\bar{H}^{\,(Q)}_a\right\rangle\nonumber\\
  &&+i\lambda\left\langle H^{(Q)}_b \sigma^{\mu\nu}F_{\mu\nu}(\rho)_{ba}\bar{H}^{\,(Q)}_a\right\rangle,
\end{eqnarray}
where both superfields $\mathcal{S}_{\mu}$ and $H^{({Q})}_a$ are defined at the heavy quark limit \cite{Wise:1992hn}. Specifically, the superfield $\mathcal{S}_{\mu}$ is composed of the $S$-wave single-charmed baryons in the $6_F$ flavor representation $\mathcal{B}_6$ with $J^P=1/2^+$ and $\mathcal{B}^*_6$ with $J^P=3/2^+$, and the $S$-wave charmed-strange mesons ${D}_s$ with $J^P=0^-$ and ${D}^{*}_s$ with $J^P=1^-$ can be constructed as the superfield $H^{({Q})}_a$, which can be expressed as \cite{Wise:1992hn,Casalbuoni:1992gi,Casalbuoni:1996pg,Yan:1992gz,Bando:1987br,Harada:2003jx,Chen:2017xat,Ding:2008gr}
\begin{eqnarray}
\mathcal{S}_{\mu}&=&-\sqrt{\frac{1}{3}}(\gamma_{\mu}+v_{\mu})\gamma^5\mathcal{B}_6+\mathcal{B}_{6\mu}^*,\\
H^{(Q)}_a&=&\frac{1+{v}\!\!\!\slash}{2}\left(D^{*(Q)\mu}_a\gamma_{\mu}-D^{(Q)}_a\gamma_5\right).
\end{eqnarray}
The corresponding conjugate fields fulfill the relations $\bar{\mathcal{S}}_{\mu}=\mathcal{S}_{\mu}^\dag\gamma^0$ and $\bar{H}^{(Q)}_a=\gamma_0H^{(Q)\dagger}_a\gamma_0$ \cite{Wise:1992hn,Casalbuoni:1992gi,Casalbuoni:1996pg,Yan:1992gz,Bando:1987br,Harada:2003jx,Chen:2017xat,Ding:2008gr}. Here, $v_{\mu}$ represents the four velocity, which is given by $v_{\mu}=(1,\bm{0})$ in the non-relativistic approximation. In addition, we also define the pseudo-Goldstone meson field $\xi$, the axial current $\mathcal{A}_\mu$, the vector current ${\cal V}_{\mu}$,  the vector meson field $\rho_{\mu}$, and the vector meson field strength tensor $F^{\mu\nu}(\rho)$, which can be represented as $\xi=e^{i\mathbb{P}/f_{\pi}}$, ${\mathcal A}_{\mu}=(\xi^{\dagger}\partial_{\mu}\xi-\xi\partial_{\mu}\xi^{\dagger})/2$, ${\mathcal V}_{\mu}=(\xi^{\dagger}\partial_{\mu}\xi+\xi\partial_{\mu}\xi^{\dagger})/2$,  $\rho_{\mu}=ig_V\mathbb{V}_{\mu}/\sqrt{2}$, and $F^{\mu\nu}(\rho)=\partial^{\mu}\rho^{\nu}-\partial^{\nu}\rho^{\mu}+\left[\rho^{\mu},\rho^{\nu}\right]$, respectively. Meanwhile, the matrices for the pseudoscalar mesons $\mathbb{P}$ and the vector mesons $\mathbb{V}_{\mu}$ are provided as follows
\begin{eqnarray}
\left.\begin{array}{l}
{\mathbb{P}} = {\left(\begin{array}{ccc}
       \frac{\pi^0}{\sqrt{2}}+\frac{\eta}{\sqrt{6}} &\pi^+ &K^+\\
       \pi^-       &-\frac{\pi^0}{\sqrt{2}}+\frac{\eta}{\sqrt{6}} &K^0\\
       K^-         &\bar K^0   &-\sqrt{\frac{2}{3}} \eta     \end{array}\right)},\\
{\mathbb{V}}_{\mu} = {\left(\begin{array}{ccc}
       \frac{\rho^0}{\sqrt{2}}+\frac{\omega}{\sqrt{2}} &\rho^+ &K^{*+}\\
       \rho^-       &-\frac{\rho^0}{\sqrt{2}}+\frac{\omega}{\sqrt{2}} &K^{*0}\\
       K^{*-}         &\bar K^{*0}   & \phi     \end{array}\right)}_{\mu},
\end{array}\right.
\end{eqnarray}
respectively. In the above effective Lagrangians, the normalization relations for the heavy hadrons $\Omega_{c}$, $\Omega_{c}^{*}$, $D_{s}$, and $D_{s}^{*}$ are denoted as \cite{Ding:2008gr}
\begin{eqnarray}
\langle 0|\Omega_{c}|css\left({1}/{2}^+\right)\rangle &=& \sqrt{2m_{\Omega_{c}}}{\left(\chi_{\frac{1}{2}m},\frac{\bm{\sigma}\cdot\bm{p}}{2m_{\Omega_{c}}}\chi_{\frac{1}{2}m}\right)^T},\nonumber\\
\langle 0|\Omega_{c}^{*\mu}|css\left({3}/{2}^+\right)\rangle &=&\sqrt{2m_{\Omega_{c}^*}}\left(\Phi_{\frac{3}{2}m}^{\mu},\frac{\bm{\sigma}\cdot\bm{p}}{2m_{\Omega_{c}^*}}\Phi_{\frac{3}{2}m}^{\mu}\right)^T,\nonumber\\
\langle 0|D_{s}|c\bar{s}\left(0^-\right)\rangle&=&\sqrt{m_{D_{s}}},\nonumber\\
\langle 0|D_{s}^{*\mu}|c\bar{s}\left(1^-\right)\rangle&=&\sqrt{m_{D_{s}^*}}\epsilon^\mu.
\end{eqnarray}
Here, the mass of the heavy hadron $i$ is denoted by $m_{i}$ ($i=\Omega_{c},\,\Omega_{c}^{*},\,D_{s},\,D_{s}^{*}$), and we define the Pauli matrix and the momentum of the charmed baryon $\Omega_{c}^{(*)}$ as $\bm{\sigma}$ and $\bm{p}$ in the normalization relations for the charmed baryon $\Omega_{c}$, $\Omega_{c}^{*}$, respectively.

The expanded effective Lagrangians among the heavy hadrons $\mathcal{B}_6^{(*)}/{D}_s^{(*)}$ and the light scalar, pseudoscalar, and vector mesons are obtained by expanding the constructed effective Lagrangians to the leading order of the pseudo-Goldstone meson field $\xi$, which can be expressed explicitly as
\begin{eqnarray}
\mathcal{L}_{\mathcal{B}_{6}^{(*)}\mathcal{B}_{6}^{(*)}f_0} &=&-l_S\langle\bar{\mathcal{B}}_6 f_0\mathcal{B}_6\rangle+l_S\langle\bar{\mathcal{B}}_{6\mu}^{*}f_0\mathcal{B}_6^{*\mu}\rangle\nonumber\\
    &&-\frac{l_S}{\sqrt{3}}\langle\bar{\mathcal{B}}_{6\mu}^{*}f_0 \left(\gamma^{\mu}+v^{\mu}\right)\gamma^5\mathcal{B}_6\rangle+h.c.,\\
\mathcal{L}_{{D}_s^{(*)}{D}_s^{(*)}f_0} &=&-2g_S{D}_a f_0 {D}_a^{\dag}+ 2g_S {D}_{a\mu}^* f_0 {D}_a^{*\mu\dag},\\
\mathcal{L}_{\mathcal{B}_6^{(*)}\mathcal{B}_6^{(*)}\mathbb{P}} &=&i\frac{g_1}{2f_{\pi}}\varepsilon^{\mu\nu\lambda\kappa}v_{\kappa}\langle\bar{\mathcal{B}}_6 \gamma_{\mu}\gamma_{\lambda}\partial_{\nu}\mathbb{P}\mathcal{B}_6\rangle\nonumber\\
    &&-i\frac{3g_1}{2f_{\pi}}\varepsilon^{\mu\nu\lambda\kappa}v_{\kappa}\langle\bar{\mathcal{B}}_{6\mu}^{*}\partial_{\nu}\mathbb{P}\mathcal{B}_{6\lambda}^*\rangle\nonumber\\
    &&+i\frac{\sqrt{3}g_1}{2f_{\pi}}v_{\kappa}\varepsilon^{\mu\nu\lambda\kappa}\langle\bar{\mathcal{B}}_{6\mu}^*\partial_{\nu}\mathbb{P}{\gamma_{\lambda}\gamma^5}\mathcal{B}_6\rangle+h.c.,\\
\mathcal{L}_{{D}_s^{(*)}{ D}_s^{(*)}\mathbb{P}}&=&-\frac{2ig}{f_{\pi}}v^{\alpha}\varepsilon_{\alpha\mu\nu\lambda}D_b^{*\mu}D_a^{*\lambda\dag}\partial^{\nu}{\mathbb{P}}_{ba}\nonumber\\
    &&-\frac{2g}{f_{\pi}}(D_b^{*\mu}D_a^{\dag}+D_bD_a^{*\mu\dag})\partial_{\mu}{\mathbb{P}}_{ba},\\
\mathcal{L}_{\mathcal{B}_6^{(*)}\mathcal{B}_6^{(*)}\mathbb{V}}&=&-\frac{\beta_Sg_V}{\sqrt{2}}\langle\bar{\mathcal{B}}_6v\cdot\mathbb{V}\mathcal{B}_6\rangle\nonumber\\
    &&-i\frac{\lambda_S g_V}{3\sqrt{2}}\langle\bar{\mathcal{B}}_6\gamma_{\mu}\gamma_{\nu}\left(\partial^{\mu}\mathbb{V}^{\nu}-\partial^{\nu}\mathbb{V}^{\mu}\right)\mathcal{B}_6\rangle\nonumber\\
    &&-\frac{\beta_Sg_V}{\sqrt{6}}\langle\bar{\mathcal{B}}_{6\mu}^*v\cdot \mathbb{V}\left(\gamma^{\mu}+v^{\mu}\right)\gamma^5\mathcal{B}_6\rangle\nonumber\\
    &&-i\frac{\lambda_Sg_V}{\sqrt{6}}\langle\bar{\mathcal{B}}_{6\mu}^*\left(\partial^{\mu}\mathbb{V}^{\nu}-\partial^{\nu}\mathbb{V}^{\mu}\right)\left(\gamma_{\nu}+v_{\nu}\right)\gamma^5\mathcal{B}_6\rangle\nonumber\\
    &&+\frac{\beta_Sg_V}{\sqrt{2}}\langle\bar{\mathcal{B}}_{6\mu}^*v\cdot {V}\mathcal{B}_6^{*\mu}\rangle\nonumber\\
    &&+i\frac{\lambda_Sg_V}{\sqrt{2}}\langle\bar{\mathcal{B}}_{6\mu}^*\left(\partial^{\mu}\mathbb{V}^{\nu}-\partial^{\nu}\mathbb{V}^{\mu}\right)\mathcal{B}_{6\nu}^*\rangle+h.c.,\\
\mathcal{L}_{{D}_s^{(*)}{D}_s^{(*)}\mathbb{V}} &=&-\sqrt{2}\beta g_V D_b D_a^{\dag} v\cdot\mathbb{V}_{ba}+\sqrt{2}\beta g_V D_{b\mu}^* D_a^{*\mu\dag}v\cdot\mathbb{V}_{ba}\nonumber\\
    &&-2\sqrt{2}i\lambda g_V D_b^{*\mu}D_a^{*\nu\dag}\left(\partial_{\mu}\mathbb{V}_{\nu}-\partial_{\nu}\mathbb{V}_{\mu}\right)_{ba}\nonumber\\
    &&-2\sqrt{2}\lambda g_V v^{\lambda}\varepsilon_{\lambda\mu\alpha\beta}(D_bD_a^{*\mu\dag}+D_b^{*\mu}D_a^{\dag})\partial^{\alpha}\mathbb{V}^{\beta}_{ba}.\nonumber\\
\end{eqnarray}
The coupling constants in the effective Lagrangians are the essential inputs for analyzing the mass spectra of the $\Omega_{c}^{(*)}{D}_s^{(*)}$-type doubly charmed molecular pentaquark candidates, which can be determined by reproducing the experimental data or calculated by using the theoretical models. From the quark model \cite{Riska:2000gd}, we can estimate the corresponding phase factors between the associated coupling constants. The following numerical calculations utilise the coupling constants as $l_S=6.20$, $g_S=0.76$, $g_1=0.94$, $g=-0.59$, $f_\pi=0.132~\rm{GeV}$, $\beta_S g_V=10.14$, $\beta g_V=-5.25$, $\lambda_S g_V=19.2~\rm{GeV}^{-1}$, and $\lambda g_V =-3.27~\rm{GeV}^{-1}$ \cite{Wang:2022mxy,Chen:2017xat,Chen:2019asm,Chen:2020kco,Wang:2020bjt,Wang:2019nwt,Chen:2018pzd,Wang:2021hql,Wang:2021ajy,Wang:2021yld,Wang:2021aql,Yang:2021sue,Wang:2020dya,Wang:2023ftp,Yalikun:2023waw}, which are frequently used to describe the interactions between the hadrons and study the masses of the observed $P_{\psi}^{N}(4312)$, $P_{\psi}^{N}(4440)$, and $P_{\psi}^{N}(4457)$ in the $\Sigma_c \bar D^{(*)}$ molecular picture \cite{Chen:2019asm}. Furthermore, we sourced the masses of the involved hadrons from the Particle Data Group \cite{ParticleDataGroup:2022pth}, which include $m_{f_0}=990.00$ MeV, $m_\eta=547.86$ MeV, $m_\phi=1019.46$ MeV, $m_{D_s}=1968.35$ MeV, $m_{D_s^*}=2112.20$ MeV, $m_{\Omega_{c}}=2695.20$ MeV, and $m_{\Omega_{c}^*}=2765.90$ MeV.

Utilizing the aforementioned discussions, the scattering amplitude $\mathcal{M}^{\Omega_{c}^{(*)}{D}_s^{(*)}\to \Omega_{c}^{(*)}{D}_s^{(*)}}(\bm{q})$ of the $\Omega_{c}^{(*)}{D}_s^{(*)}\to \Omega_{c}^{(*)}{D}_s^{(*)}$ process can be obtained. By using the Breit approximation and the non-relativistic normalization \cite{Breitapproximation}, the effective potential in the momentum space $\mathcal{V}_E^{\Omega_{c}^{(*)}{D}_s^{(*)}\to \Omega_{c}^{(*)}{D}_s^{(*)}}(\bm{q})$ can be obtained, and we can write the specific expression explicitly as \cite{Breitapproximation}
\begin{eqnarray}
\mathcal{V}_E^{\Omega_{c}^{(*)}{D}_s^{(*)}\to \Omega_{c}^{(*)}{D}_s^{(*)}}(\bm{q})=-\frac{\mathcal{M}^{\Omega_{c}^{(*)}{D}_s^{(*)}\to \Omega_{c}^{(*)}{D}_s^{(*)}}(\bm{q})} {\sqrt{2m_{\Omega_{c}^{(*)}}2m_{{D}_s^{(*)}}2m_{\Omega_{c}^{(*)}}2m_{{D}_s^{(*)}}}}.
\end{eqnarray}
Finally, it is essential to obtain the effective potential in the coordinate space $\mathcal{V}_E^{\Omega_{c}^{(*)}{D}_s^{(*)}\to \Omega_{c}^{(*)}{D}_s^{(*)}}(\bm{r})$, and this is due to the fact that our present work utilizes the coupled channel Schr\"{o}dinger equation solved in the coordinate space to analyze the bound state properties of the $\Omega_{c}^{(*)} D_s^{(*)}$ systems. In the specific calculations, the effective potential in the coordinate space $\mathcal{V}_E^{\Omega_{c}^{(*)}{D}_s^{(*)}\to \Omega_{c}^{(*)}{D}_s^{(*)}}(\bm{r})$ can be obtained by performing the Fourier transformation for the effective potential in the momentum space $\mathcal{V}_E^{\Omega_{c}^{(*)}{D}_s^{(*)}\to \Omega_{c}^{(*)}{D}_s^{(*)}}(\bm{q})$ and the form factor $\mathcal{F}(q^2,m_{\mathcal{E}}^2)$, which is expressed as
\begin{eqnarray}
&&\mathcal{V}_E^{\Omega_{c}^{(*)}{D}_s^{(*)}\to \Omega_{c}^{(*)}{D}_s^{(*)}}(\bm{r})\nonumber\\ &&=\int\frac{d^3\bm{q}}{(2\pi)^3}e^{i\bm{q}\cdot\bm{r}}\mathcal{V}_E^{\Omega_{c}^{(*)}{D}_s^{(*)}\to \Omega_{c}^{(*)}{D}_s^{(*)}}(\bm{q})\mathcal{F}^2(q^2,m_{\mathcal{E}}^2).
\end{eqnarray}
It is well known that our discussed baryons and mesons exist the complex inner structures, and the form factor is commonly utilized to represent the off-shell effect of the exchanged light mesons and the finite size effect of the discussed hadrons. Thus, the monopole-type form factor $\mathcal{F}(q^2,m_{\mathcal{E}}^2) = {(\Lambda^2-m_{\mathcal{E}}^2)}/{(\Lambda^2-q^2)}$ is introduced into both interactive vertex $\Gamma^{h_1h_3\mathcal{E}}_{(\mu)}$ and $\Gamma^{h_2h_4\mathcal{E}}_{(\nu)}$ during the aforementioned Fourier transformation  \cite{Tornqvist:1993ng,Tornqvist:1993vu}, where the introduced cutoff parameter, the mass, and the four momentum of the exchanged light meson are defined as $\Lambda$, $m_{\mathcal{E}}$, and $q$, respectively.

The above procedure can be used to derive the OBE effective potentials in the coordinate space for the $\Omega_{c}^{(*)}{D}_s^{(*)}$ systems. In Table~\ref{potentials}, we summarise the obtained OBE effective potentials in the coordinate space for the $\Omega_{c}^{(*)}{D}_s^{(*)}$ systems, which include the OBE effective potentials and the operators.

\renewcommand\tabcolsep{0.42cm}
\renewcommand{\arraystretch}{1.50}
\begin{table*}[!htbp]
\caption{The obtained OBE effective potentials in the coordinate space for the $\Omega_{c}^{(*)}{D}_s^{(*)}$ systems. The tensor force operator, denoted as $T({\bm x},{\bm y})$, can be defined as $3\left(\hat{\bm r} \cdot {\bm x}\right)\left(\hat{\bm r} \cdot {\bm y}\right)-{\bm x} \cdot {\bm y}$, where $\hat{\bm r}$ is the radial unit vector.}\label{potentials}
\begin{tabular}{ccc}\toprule[1pt]\toprule[1pt]
\multicolumn{3}{c}{OBE effective potentials}\\\midrule[1.0pt]
\multicolumn{3}{l}{$\mathcal{V}_E^{\Omega_{c}D_s\rightarrow\Omega_{c} D_s}(\bm{r})$=$-l_Sg_SY_{f_0}+\frac{\beta_S \beta g_{V}^2}{2}Y_{\phi}$}\\
\multicolumn{3}{l}{$\mathcal{V}_E^{\Omega_{c}^*D_s\rightarrow\Omega_{c}^*D_s}(\bm{r})$=$-l_Sg_S\mathcal{O}_{1}Y_{f_0}+\frac{\beta_S \beta g_{V}^2}{2}\mathcal{O}_{1}Y_{\phi}$}\\
\multicolumn{3}{l}{$\mathcal{V}_E^{\Omega_{c}D_s^*\rightarrow\Omega_{c} D_s^*}(\bm{r})$=$-l_Sg_S\mathcal{O}_{2}Y_{f_0}+\frac{2}{9}\frac{g_1 g}{f_\pi^2}\left(\mathcal{O}_{3}\mathcal{O}_r+\mathcal{O}_{4}\mathcal{P}_r\right)Y_{\eta}+\frac{\beta_S \beta g_{V}^2}{2}\mathcal{O}_{2}Y_{\phi}+\frac{2\lambda_S \lambda g_V^2}{9}\left(2\mathcal{O}_{3}\mathcal{O}_r-\mathcal{O}_{4}\mathcal{P}_r\right)Y_{\phi}$}\\
\multicolumn{3}{l}{$\mathcal{V}_E^{\Omega_{c}^* D_s^*\rightarrow\Omega_{c}^{*} D_s^*}(\bm{r})$=$-l_Sg_S\mathcal{O}_{5}Y_{f_0}-\frac{1}{3}\frac{g_1 g}{f_\pi^2}\left(\mathcal{O}_{6}\mathcal{O}_r+\mathcal{O}_{7}\mathcal{P}_r\right)Y_{\eta}+\frac{\beta_S \beta g_{V}^2}{2}\mathcal{O}_{5}Y_{\phi}-\frac{\lambda_S \lambda g_V^2}{3}\left(2\mathcal{O}_{6}\mathcal{O}_r-\mathcal{O}_{7}\mathcal{P}_r\right)Y_{\phi}$}\\
\multicolumn{3}{l}{$\mathcal{V}_E^{\Omega_{c} D_s\rightarrow\Omega_{c}^* D_s}(\bm{r})$=$\frac{l_Sg_S}{\sqrt{3}}\mathcal{O}_8Y_{f_01}-\frac{\beta_S \beta g_{V}^2}{2\sqrt{3}}\mathcal{O}_8Y_{\phi1}$}\\
\multicolumn{3}{l}{$\mathcal{V}_E^{\Omega_{c} D_s\rightarrow\Omega_{c} D_s^*}(\bm{r})$=$-\frac{2}{9}\frac{g_1 g}{f_\pi^2}\left(\mathcal{O}_9\mathcal{O}_r+\mathcal{O}_{10}\mathcal{P}_r\right)Y_{\eta2}+\frac{2\lambda_S \lambda g_V^2}{9}\left(2\mathcal{O}_9\mathcal{O}_r-\mathcal{O}_{10}\mathcal{P}_r\right)Y_{\phi2}$}\\
\multicolumn{3}{l}{$\mathcal{V}_E^{\Omega_{c} D_s\rightarrow\Omega_{c}^* D_s^*}(\bm{r})$=$\frac{1}{3\sqrt{3}}\frac{g_1 g}{f_\pi^2}\left(\mathcal{O}_{11}\mathcal{O}_r+\mathcal{O}_{12}\mathcal{P}_r\right)Y_{\eta3}-\frac{\lambda_S \lambda g_V^2}{3\sqrt{3}}\left(2\mathcal{O}_{11}\mathcal{O}_r-\mathcal{O}_{12}\mathcal{P}_r\right)Y_{\phi3}$}\\
\multicolumn{3}{l}{$\mathcal{V}_E^{\Omega_{c}^* D_s\rightarrow\Omega_{c} D_s^*}(\bm{r})$=$\frac{1}{3\sqrt{3}}\frac{g_1 g}{f_\pi^2}\left(\mathcal{O}_{13}\mathcal{O}_r+\mathcal{O}_{14}\mathcal{P}_r\right)Y_{\eta4}-\frac{\lambda_S \lambda g_V^2}{3\sqrt{3}}\left(2\mathcal{O}_{13}\mathcal{O}_r-\mathcal{O}_{14}\mathcal{P}_r\right)Y_{\phi4}$}\\
\multicolumn{3}{l}{$\mathcal{V}_E^{\Omega_{c}^* D_s\rightarrow\Omega_{c}^* D_s^*}(\bm{r})$=$-\frac{1}{3}\frac{g_1 g}{f_\pi^2}\left(\mathcal{O}_{15}\mathcal{O}_r+\mathcal{O}_{16}\mathcal{P}_r\right)Y_{\eta5}+\frac{\lambda_S \lambda g_V^2}{3}\left(2\mathcal{O}_{15}\mathcal{O}_r-\mathcal{O}_{16}\mathcal{P}_r\right)Y_{\phi5}$}\\
\multicolumn{3}{l}{$\mathcal{V}_E^{\Omega_{c} D_s^*\rightarrow\Omega_{c}^* D_s^*}(\bm{r})$=$\frac{l_Sg_S}{\sqrt{3}}\mathcal{O}_{17}Y_{f_06}+\frac{1}{3\sqrt{3}}\frac{g_1 g}{f_\pi^2}\left(\mathcal{O}_{18}\mathcal{O}_r+\mathcal{O}_{19}\mathcal{P}_r\right)Y_{\eta6}-\frac{\beta_S \beta g_{V}^2}{2\sqrt{3}}\mathcal{O}_{17}Y_{\phi6}+\frac{\lambda_S \lambda g_V^2}{3\sqrt{3}}\left(2\mathcal{O}_{18}\mathcal{O}_r-\mathcal{O}_{19}\mathcal{P}_r\right)Y_{\phi6}$}\\
\multicolumn{3}{l}{$\mathcal{O}_r = \frac{1}{r^2}\frac{\partial}{\partial r}r^2\frac{\partial}{\partial r}$~~~~$\mathcal{P}_r = r\frac{\partial}{\partial r}\frac{1}{r}\frac{\partial}{\partial r}$~~~~$Y_{\mathcal{E}i}= \dfrac{e^{-m_{\mathcal{E}i}r}-e^{-\Lambda_ir}}{4\pi r}-\dfrac{\Lambda_i^2-m_{\mathcal{E}i}^2}{8\pi\Lambda_i}e^{-\Lambda_ir}$~~~~$m_{\mathcal{E}i}=\sqrt{m_{\mathcal{E}}^2-q_i^2}$~~~~$\Lambda_i=\sqrt{\Lambda^2-q_i^2}$}\\
\multicolumn{3}{l}{$q_1=0.04$ GeV~~~~$q_2=0.06$ GeV~~~~$q_3=0.02$ GeV~~~~$q_4=0.10$ GeV~~~~$q_5=0.06$ GeV~~~~$q_6=0.04$ GeV}\\\midrule[1.0pt]
\multicolumn{3}{c}{Operators}\\\midrule[1.0pt]
\multicolumn{3}{l}{$\mathcal{O}_{1}=\sum_{a,b,m,n}C^{\frac{3}{2},a+b}_{\frac{1}{2}a,1b}C^{\frac{3}{2},m+n}_{\frac{1}{2}m,1n}\chi^{\dagger a}_{3}\left({\bm\epsilon^{\dagger b}_{3}}\cdot{\bm\epsilon^{n}_{1}}\right)\chi^{m}_1$~~~~~~
$\mathcal{O}_{2}=\chi^{\dagger}_3\left({\bm\epsilon^{\dagger}_{4}}\cdot{\bm\epsilon_{2}}\right)\chi_1$~~~~~~
$\mathcal{O}_{3}=\chi^{\dagger}_3\left[{\bm\sigma}\cdot\left(i{\bm\epsilon_{2}}\times{\bm\epsilon^{\dagger}_{4}}\right)\right]\chi_1$~~~~~~
$\mathcal{O}_{4}=\chi^{\dagger}_3T({\bm\sigma},i{\bm\epsilon_{2}}\times{\bm\epsilon^{\dagger}_{4}})\chi_1$}\\
\multicolumn{3}{l}{$\mathcal{O}_{5}=\sum_{a,b,m,n}C^{\frac{3}{2},a+b}_{\frac{1}{2}a,1b}C^{\frac{3}{2},m+n}_{\frac{1}{2}m,1n}\chi^{\dagger a}_3\left({\bm\epsilon^{n}_{1}}\cdot{\bm\epsilon^{\dagger b}_{3}}\right)\left({\bm\epsilon_{2}}\cdot{\bm\epsilon^{\dagger}_{4}}\right)\chi^m_1$~~~~~~
$\mathcal{O}_{6}=\sum_{a,b,m,n}C^{\frac{3}{2},a+b}_{\frac{1}{2}a,1b}C^{\frac{3}{2},m+n}_{\frac{1}{2}m,1n}\chi^{\dagger a}_3\left({\bm\epsilon^{n}_{1}}\times{\bm\epsilon^{\dagger b}_{3}}\right)\cdot\left({\bm\epsilon_{2}}\times{\bm\epsilon^{\dagger}_{4}}\right)\chi^m_1$}\\
\multicolumn{3}{l}{$\mathcal{O}_{7}=\sum_{a,b,m,n}C^{\frac{3}{2},a+b}_{\frac{1}{2}a,1b}C^{\frac{3}{2},m+n}_{\frac{1}{2}m,1n}\chi^{\dagger a}_3T({\bm\epsilon^{n}_{1}}\times{\bm\epsilon^{\dagger b}_{3}},{\bm\epsilon_{2}}\times{\bm\epsilon^{\dagger}_{4}})\chi^m_1$~~~~~~
$\mathcal{O}_{8}=\sum_{a,b}C^{\frac{3}{2},a+b}_{\frac{1}{2}a,1b}\chi^{\dagger a}_{3}\left({\bm\epsilon^{\dagger b}_{3}}\cdot{\bm\sigma}\right)\chi_1$~~~~~~
$\mathcal{O}_{9}=\chi^{\dagger}_3\left({\bm\sigma}\cdot{\bm\epsilon^{\dagger}_{4}}\right)\chi_1$~~~~~~
$\mathcal{O}_{10}=\chi^{\dagger}_3T({\bm\sigma},{\bm\epsilon^{\dagger}_{4}})\chi_1$}\\
\multicolumn{3}{l}{$\mathcal{O}_{11}=\sum_{a,b}C^{\frac{3}{2},a+b}_{\frac{1}{2}a,1b}\chi^{\dagger a}_3\left[{\bm\epsilon^{\dagger}_{4}}\cdot\left(i{\bm\sigma}\times{\bm\epsilon^{\dagger b}_{3}}\right)\right]\chi_1$~~~~~~
$\mathcal{O}_{12}=\sum_{a,b}C^{\frac{3}{2},a+b}_{\frac{1}{2}a,1b}\chi^{\dagger a}_3T({\bm\epsilon^{\dagger}_{4}},i{\bm\sigma}\times{\bm\epsilon^{\dagger b}_{3}})\chi_1$~~~~~~
$\mathcal{O}_{13}=\sum_{a,b}C^{\frac{3}{2},a+b}_{\frac{1}{2}a,1b}\chi^{\dagger}_3\left[{\bm\epsilon^{\dagger}_{4}}\cdot\left(i{\bm\sigma}\times{\bm\epsilon^{b}_{1}}\right)\right]\chi^a_1$}\\
\multicolumn{3}{l}{$\mathcal{O}_{14}=\sum_{a,b}C^{\frac{3}{2},a+b}_{\frac{1}{2}a,1b}\chi^{\dagger}_3T({\bm\epsilon^{\dagger}_{4}},i{\bm\sigma}\times{\bm\epsilon^{b}_{1}})\chi^a_1$~~~~~~
$\mathcal{O}_{15}=\sum_{a,b,m,n}C^{\frac{3}{2},a+b}_{\frac{1}{2}a,1b}C^{\frac{3}{2},m+n}_{\frac{1}{2}m,1n}\chi^{\dagger a}_3\left[{\bm\epsilon^{\dagger}_{4}}\cdot\left(i{\bm\epsilon^n_{1}}\times{\bm\epsilon^{\dagger b}_{3}}\right)\right]\chi^m_1$}\\
\multicolumn{3}{l}{$\mathcal{O}_{16}=\sum_{a,b,m,n}C^{\frac{3}{2},a+b}_{\frac{1}{2}a,1b}C^{\frac{3}{2},m+n}_{\frac{1}{2}m,1n}\chi^{\dagger a}_3T({\bm\epsilon^{\dagger}_{4}},i{\bm\epsilon^n_{1}}\times{\bm\epsilon^{\dagger b}_{3}})\chi^m_1$~~~~~~
$\mathcal{O}_{17}=\sum_{a,b}C^{\frac{3}{2},a+b}_{\frac{1}{2}a,1b}\chi^{\dagger a}_3\left({\bm\sigma}\cdot{\bm\epsilon^{\dagger b}_{3}}\right)\left({\bm\epsilon_{2}}\cdot{\bm\epsilon^{\dagger}_{4}}\right)\chi_1$}\\
\multicolumn{3}{l}{$\mathcal{O}_{18}=\sum_{a,b}C^{\frac{3}{2},a+b}_{\frac{1}{2}a,1b}\chi^{\dagger a}_3\left({\bm\sigma}\times{\bm\epsilon^{\dagger b}_{3}}\right)\cdot\left({\bm\epsilon_{2}}\times{\bm\epsilon^{\dagger}_{4}}\right)\chi_1$~~~~~~
$\mathcal{O}_{19}=\sum_{a,b}C^{\frac{3}{2},a+b}_{\frac{1}{2}a,1b}\chi^{\dagger a}_3T({\bm\sigma}\times{\bm\epsilon^{\dagger b}_{3}},{\bm\epsilon_{2}}\times{\bm\epsilon^{\dagger}_{4}})\chi_1$}\\
\multicolumn{3}{l}{$\mathcal{O}_{1}^{J=3/2}$=diag(1,1)~~~~~~$\mathcal{O}_{2}^{J=1/2}$=diag(1,1)~~~~~~$\mathcal{O}_{2}^{J=3/2}$=diag(1,1,1)~~~~~~$\mathcal{O}_{3}^{J=1/2}$=diag($-2$,$1$)~~~~~~$\mathcal{O}_{3}^{J=3/2}$=diag($1$,$-2$,$1$)~~~~~~$\mathcal{O}_{4}^{J=1/2}$=$\left(\begin{array}{cc} 0 & -\sqrt{2} \\ -\sqrt{2} & -2\end{array}\right)$}\\
\multicolumn{3}{l}{$\mathcal{O}_{4}^{J=3/2}$=$\left(\begin{array}{ccc} 0 & 1& 2 \\ 1 & 0& -1 \\ 2 & -1& 0 \end{array}\right)$~~~~~~$\begin{array}{l} \mathcal{O}_{5}^{J=1/2}={\rm diag}(1,1,1)\\ \mathcal{O}_{6}^{J=1/2}={\rm diag}(\frac{5}{3},\frac{2}{3},-1) \end{array}$~~~~~~$\begin{array}{l}
\mathcal{O}_{5}^{J=3/2}={\rm diag}(1,1,1,1)\\ \mathcal{O}_{6}^{J=3/2}={\rm diag}(\frac{2}{3},\frac{5}{3},\frac{2}{3},-1) \end{array}$~~~~~~$\begin{array}{l}\mathcal{O}_{5}^{J=5/2}={\rm diag}(1,1,1,1)\\ \mathcal{O}_{6}^{J=5/2}={\rm diag}(-1,\frac{5}{3},\frac{2}{3},-1)
\end{array}$~~~~~~}\\
\multicolumn{3}{l}{$\mathcal{O}_{7}^{J=1/2}$=$\left(\begin{array}{ccc} 0 & -\frac{7}{3\sqrt{5}}& \frac{2}{\sqrt{5}} \\ -\frac{7}{3\sqrt{5}} & \frac{16}{15}& -\frac{1}{5} \\ \frac{2}{\sqrt{5}} &-\frac{1}{5}& \frac{8}{5} \end{array}\right)$
           ~~~~~~$\mathcal{O}_{7}^{J=3/2}$=$\left(\begin{array}{cccc} 0 & \frac{7}{3\sqrt{10}}& -\frac{16}{15}& -\frac{\sqrt{7}}{5\sqrt{2}}\\ \frac{7}{3\sqrt{10}} & 0& -\frac{7}{3\sqrt{10}} & -\frac{2}{\sqrt{35}} \\ -\frac{16}{15} & -\frac{7}{3\sqrt{10}}& 0& -\frac{1}{\sqrt{14}} \\-\frac{\sqrt{7}}{5\sqrt{2}}&-\frac{2}{\sqrt{35}} &-\frac{1}{\sqrt{14}}&\frac{4}{7}\end{array}\right)$
           ~~~~~~$\mathcal{O}_{7}^{J=5/2}$=$\left(\begin{array}{cccc} 0 & \frac{2}{\sqrt{15}}& \frac{\sqrt{7}}{5\sqrt{3}}& -\frac{2\sqrt{14}}{5}\\ \frac{2}{\sqrt{15}} & 0& \frac{\sqrt{7}}{3\sqrt{5}} & -\frac{4\sqrt{2}}{\sqrt{105}} \\ \frac{\sqrt{7}}{5\sqrt{3}} & \frac{\sqrt{7}}{3\sqrt{5}}& -\frac{16}{21}& -\frac{\sqrt{2}}{7\sqrt{3}} \\-\frac{2\sqrt{14}}{5}&-\frac{4\sqrt{2}}{\sqrt{105}} &-\frac{\sqrt{2}}{7\sqrt{3}}&-\frac{4}{7}\end{array}\right)$}\\
\multicolumn{3}{l}{$\mathcal{O}_{9}^{J=1/2}=\sqrt{3}$~~~~~~$\mathcal{O}_{11}^{J=1/2}=\sqrt{2}$~~~~~~$\mathcal{O}_{18}^{J=1/2}=-\sqrt{{2}/{3}}$~~~~~~$\mathcal{O}_{i}^{J=1/2}=0\,(i = 10, \,12, \,17, \,19)$}\\
\multicolumn{3}{l}{$\mathcal{O}_{13}^{J=3/2}=1$~~~~~~$\mathcal{O}_{15}^{J=3/2}=\sqrt{{5}/{3}}$~~~~~~$\mathcal{O}_{18}^{J=3/2}=-\sqrt{{5}/{3}}$~~~~~~$\mathcal{O}_{i}^{J=3/2}=0\,(i = 14, \,16, \,17, \,19)$}\\
\bottomrule[1pt]\bottomrule[1pt]
\end{tabular}
\end{table*}

\subsection{The prediction of mass spectra of the $\Omega_{c}^{(*)}{D}_s^{(*)}$ molecules}

Based on the obtained OBE effective potentials in the coordinate space for the $\Omega_{c}^{(*)}{D}_s^{(*)}$ systems (see Table~\ref{potentials}), we can obtain the binding energy $E$ and the corresponding spatial wave functions $\phi_i(r)$ by solving the coupled channel Schr\"{o}dinger equation. Furthermore, the root-mean-square radius $r_{\rm RMS}$ and the probabilities of the individual channel $P_i$ can be calculated from the obtained spatial wave functions. These physical quantities can provide the useful hints for predicting the mass spectra of the $\Omega_{c}^{(*)}{D}_s^{(*)}$-type doubly charmed molecular pentaquark candidates.

When discussing the hadronic molecular candidates in the context of the OBE model, it is necessary to make a few remarks before presenting the numerical calculations. (i) In the OBE model,  the free parameter is the cutoff $\Lambda$ in the monopole-type form factor to study the mass spectra of the $\Omega_{c}^{(*)}{D}_s^{(*)}$-type doubly charmed molecular pentaquark candidates, and we aim to search for the loosely bound state solutions for the $\Omega_{c}^{(*)}{D}_s^{(*)}$ systems by adjusting the cutoff parameter between 1.0 and 2.5 GeV in our numerical calculations. Based on the results of studying the bound state properties of the deuteron, a loosely bound state is more likely to be considered as the hadronic molecular candidate with the cutoff parameter in the monopole-type form factor around 1.0 GeV \cite{Machleidt:1987hj,Epelbaum:2008ga,Esposito:2014rxa,Chen:2016qju,Tornqvist:1993ng,Tornqvist:1993vu,Wang:2019nwt,Chen:2017jjn}. Moreover, the study of the masses of the observed $P_{\psi}^{N}$ \cite{Aaij:2019vzc}, $P_{\psi s}^{\Lambda}$ \cite{LHCb:2020jpq,LHCb:2022ogu}, and $T_{cc}$ \cite{LHCb:2021vvq} in the hadronic molecular picture also suggests that the cutoff parameter in the monopole-type form factor around 1.0 GeV is an appropriate choice to study the hadronic molecular candidates. (ii) For the ideal hadronic molecular candidate, the binding energy is expected to be at most tens of MeV, which can ensure that there is no significant overlap for the two constituent hadrons in the spatial distribution  \cite{Chen:2016qju,Chen:2017xat}. (iii) Given that the observed $P_{\psi}^{N}$ \cite{Aaij:2019vzc}, $P_{\psi s}^{\Lambda}$ \cite{LHCb:2020jpq,LHCb:2022ogu}, and $T_{cc}$ \cite{LHCb:2021vvq} can be explained as the $S$-wave hadronic molecular candidates, this study predominantly concentrates on the $S$-wave $\Omega_{c}^{(*)}{D}_s^{(*)}$ systems.

Now, we analyse the bound state properties of the $\Omega_{c}{D}_s$, $\Omega_{c}^{*}{D}_s$, $\Omega_{c}{D}_s^{*}$, and $\Omega_{c}^{*}{D}_s^{*}$ systems by solving the coupled channel Schr\"{o}dinger equation in the context of the single channel analysis, the $S$-$D$ wave mixing analysis, and the coupled channel analysis, which allow us to predict the mass spectra of the $\Omega_{c}^{(*)}{D}_s^{(*)}$-type doubly charmed molecular pentaquark candidates.

\subsubsection{Bound state properties of the $\Omega_{c}{D}_s$ system}

\begin{figure}[htbp]
\centering
\includegraphics[width=7.5cm]{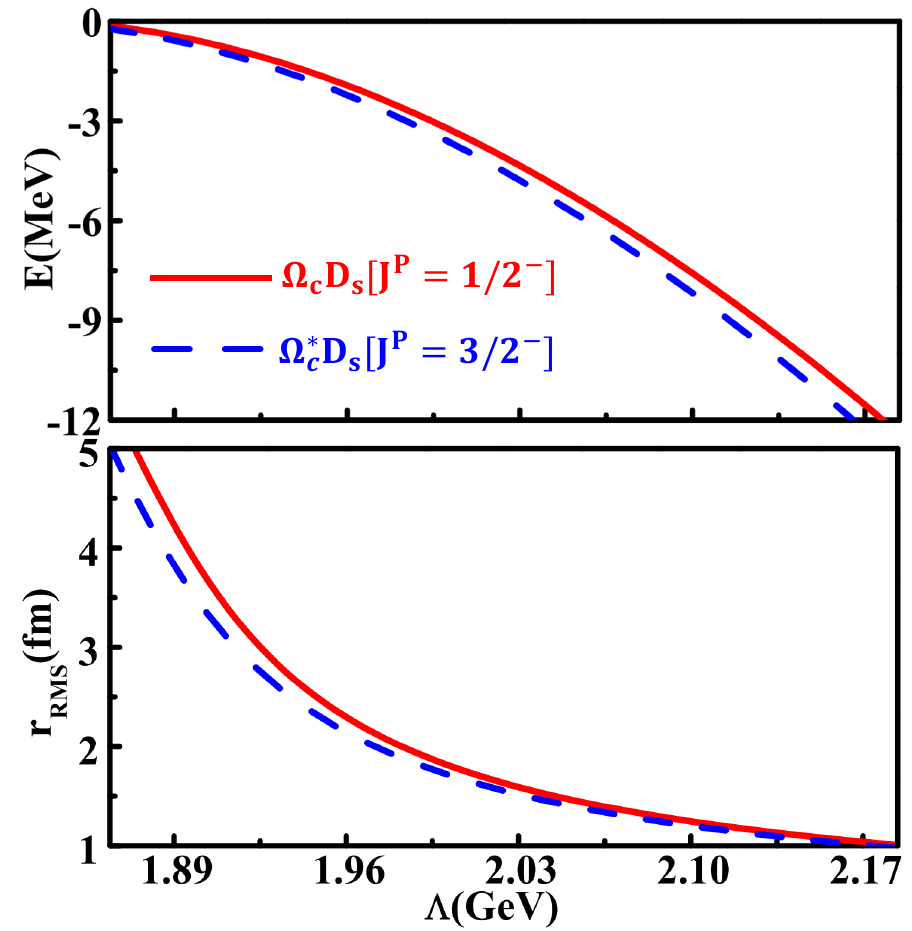}
\caption{Bound state properties for the $\Omega_{c}{D}_s$ state with $J^P={1}/{2}^{-}$ and the $\Omega_{c}^*{D}_s$ state with $J^P={3}/{2}^{-}$.}\label{massspectra1}
\end{figure}

The $\Omega_{c}{D}_s$ system has the attractive forces provided by the $f_0$ and $\phi$ exchange potentials within the OBE model, and Figure \ref{massspectra1} displays the obtained bound state properties for the $\Omega_{c}{D}_s$ state with $J^P={1}/{2}^{-}$.

The single channel analysis shows that the loosely bound state solutions for the $\Omega_{c}{D}_s$ state with $J^P={1}/{2}^{-}$ exist at the cutoff value $\Lambda$ greater than 1.88 GeV. Thus, the $\Omega_{c}{D}_s$ state with $J^P={1}/{2}^{-}$ can be proposed as the possible candidate of the doubly charmed molecular pentaquark. Moreover, we can incorporate the coupled channel effect involving the $\Omega_{c}{D}_s$, $\Omega_{c}{D}_s^{*}$, and $\Omega_{c}^{*}{D}_s^{*}$ channels to analyze the bound state properties for the $\Omega_{c}{D}_s$ state with $J^P={1}/{2}^{-}$. After performing the numerical calculations, we find that the $\Omega_{c}{D}_s$ system with the lowest mass threshold is not the dominant channel, which makes the corresponding size of this coupled system is less than 0.4 fm and it contradicts the idea of the loose molecular picture \cite{Chen:2016qju,Chen:2017xat}. In the present work, we do not list such numerical results when considering the coupled channel effect. In fact, this phenomenon has already been explored in previous works \cite{Wang:2023ftp,Wang:2022mxy}. When employing the same cutoff value for the coupled channels $\Omega_{c}{D}_s$, $\Omega_{c}{D}_s^{*}$, and $\Omega_{c}^{*}{D}_s^{*}$, the effective interaction of the $\Omega_{c}{D}_s$ channel with $J^P={1}/{2}^{-}$ is much weaker than those of other coupled channels with $J^P={1}/{2}^{-}$, which can lead to the dominant channel not being the $\Omega_{c}{D}_s$ system. In the future studies, we can further discuss the bound state properties for the $\Omega_{c}{D}_s$ state with $J^P={1}/{2}^{-}$ by considering the coupled channel effect and taking different cutoff values for the involved coupled channels once the relevant experimental data become available.

\subsubsection{Bound state properties of the $\Omega_{c}^*{D}_s$ system}

The total effective potential of the $\Omega_{c}^*{D}_s$ system comprises contribution from the $f_0$ and $\phi$ exchanges within the OBE model. Moreover, there is no contribution of the tensor force from the $S$-$D$ wave mixing effect (see Table~\ref{potentials}). Hence, the $\Omega_{c}^*{D}_s$ state with $J^P={3}/{2}^{-}$ has identical bound state properties with and without considering the $S$-$D$ wave mixing effect, and the contribution of the $|{}^4\mathbb{D}_{{3}/{2}}\rangle$ channel is zero.

In Figure \ref{massspectra1}, we present the obtained bound state properties for the $\Omega_{c}^*{D}_s$ state with $J^P={3}/{2}^{-}$. For the $\Omega_{c}^*{D}_s$ state with $J^P={3}/{2}^{-}$, we obtain the bound state solutions by choosing the cutoff value $\Lambda$ around 1.87 GeV or even larger in the single channel and $S$-$D$ wave mixing analysis. According to the above analysis, the $\Omega_{c}^*{D}_s$ state with $J^P={3}/{2}^{-}$ can be assigned as the possible candidate of the doubly charmed molecular pentaquark. In addition, the bound state properties of the $\Omega_{c}^*{D}_s$ state with $J^P={3}/{2}^{-}$ can also be discussed by conducting the coupled channel analysis involving the $\Omega_{c}^{*}{D}_s$, $\Omega_{c}{D}_s^{*}$, and $\Omega_{c}^{*}{D}_s^{*}$ channels. Similar to the case of the $\Omega_{c}{D}_s$ state with $J^P={1}/{2}^{-}$, our numerical results indicate that the primary channel is not the $\Omega_{c}^*{D}_s$ system when the coupled channel effect is taken into account. This can be attributed to the fact that the effective potential of the $\Omega_{c}^{*}{D}_s$ channel with $J^P={3}/{2}^{-}$ is significantly weaker compared to the effective potentials of the other coupled channels with $J^P={3}/{2}^{-}$ if using the same cutoff value for the involved coupled channels.

\subsubsection{Bound state properties of the $\Omega_{c}{D}_s^*$ system}

The $\Omega_{c}{D}_s^*$ system has the $\eta$, $f_0$, and $\phi$ exchange potentials within the OBE model, and its spin-parity quantum numbers $J^P$ can be either $1/2^-$ or $3/2^-$. In our concrete calculations, we first search for the loosely bound state solutions for the $\Omega_{c}{D}_s^*$ system by performing the single channel analysis. And then, we discuss the bound state properties for the $\Omega_{c}{D}_s^*$ system by incorporating the contribution from the $S$-$D$ wave mixing effect and the coupled channel effect. In Table \ref{massspectra3}, we present the bound state properties for the $\Omega_{c}{D}_s^*$ system obtained in three ways: the single channel analysis, the $S$-$D$ wave mixing analysis, and the coupled channel analysis, respectively.

\renewcommand\tabcolsep{0.20cm}
\renewcommand{\arraystretch}{1.50}
\begin{table}[!htbp]
\caption{The obtained bound state properties for the $\Omega_{c}{D}_s^*$ system. The units of the cutoff parameter $\Lambda$, the binding energy $E$, and the root-mean-square radius $r_{\rm RMS}$ are $ \rm{GeV}$, $\rm {MeV}$, and $\rm {fm}$, respectively. The dominant probabilities for the corresponding channels are displayed in bold font. Here, the obtained results labeled as Case I, Case II, and Case III correspond to the single channel analysis, the $S$-$D$ wave mixing analysis, and the coupled channel analysis, respectively.}\label{massspectra3}
\begin{tabular}{c|c|cccc}\toprule[1pt]\toprule[1pt]
\multirow{12}{*}{$J^P=\frac{1}{2}^{-}$}&\multirow{4}{*}{Case I}&$\Lambda$ &$E$  &$r_{\rm RMS}$ &\\
\cline{3-6}
&&1.32&$-0.54$ &3.75&  \\
&&1.34&$-3.68$ &1.57&      \\
&&1.36&$-9.42$ &1.02&      \\
\cline{2-6}
&\multirow{4}{*}{Case II}&$\Lambda$ &$E$  &$r_{\rm RMS}$ &P(${}^2\mathbb{S}_{\frac{1}{2}}/{}^4\mathbb{D}_{\frac{1}{2}})$\\
\cline{3-6}
&&1.32&$-0.76$ &3.72&\textbf{99.92}/0.08\\
&&1.34&$-4.11$ &1.49&\textbf{99.88}/0.12\\
&&1.36&$-9.96$ &1.00&\textbf{99.87}/0.13\\
\cline{2-6}
&\multirow{4}{*}{Case III}&$\Lambda$ &$E$  &$r_{\rm RMS}$ &P($\Omega_{c} D_s^*/\Omega_{c}^{*} D_s^*$)\\
\cline{3-6}
&&1.29&$-1.20$ &2.60&\textbf{95.64}/4.36\\
&&1.30&$-3.87$ &1.47&\textbf{91.35}/8.65\\
&&1.31&$-8.03$ &1.03&\textbf{86.51}/13.49\\\hline
\multirow{12}{*}{$J^P=\frac{3}{2}^{-}$}&\multirow{4}{*}{Case I}&$\Lambda$ &$E$  &$r_{\rm RMS}$ &\\
\cline{3-6}
&&2.34&$-0.28$ &4.90&  \\
&&2.42&$-0.74$ &3.56&      \\
&&2.50&$-1.41$ &2.73&      \\
\cline{2-6}
&\multirow{4}{*}{Case II}&$\Lambda$ &$E$  &$r_{\rm RMS}$ &P(${}^4\mathbb{S}_{\frac{3}{2}}/{}^2\mathbb{D}_{\frac{3}{2}}/{}^4\mathbb{D}_{\frac{3}{2}})$\\
\cline{3-6}
&&2.31&$-0.24$ &5.06&\textbf{99.96}/0.01/0.03\\
&&2.41&$-0.89$ &3.32&\textbf{99.94}/0.01/0.05\\
&&2.50&$-1.83$ &2.44&\textbf{99.92}/0.02/0.06\\
\cline{2-6}
&\multirow{4}{*}{Case III}&$\Lambda$ &$E$  &$r_{\rm RMS}$ &P($\Omega_{c} D_s^*/\Omega_{c}^{*} D_s^*$)\\
\cline{3-6}
&&1.50&$-1.08$ &2.62&\textbf{79.66}/20.34\\
&&1.51&$-4.50$ &1.21&\textbf{61.10}/38.90\\
&&1.52&$-9.70$ &0.80&48.91/\textbf{51.09}\\
\bottomrule[1pt]\bottomrule[1pt]
\end{tabular}
\end{table}

For the $\Omega_{c}{D}_s^*$ state with $J^P={1}/{2}^{-}$, we can obtain the bound state solutions by choosing the cutoff value $\Lambda$ around 1.32 GeV and considering only the contribution of the $S$-wave channel. And then, the bound state solutions can be found with the cutoff value $\Lambda$ around 1.32 GeV by including the contribution of the $D$-wave channel. However, the $S$-$D$ wave mixing effect has a minor role in generating the $\Omega_{c}{D}_s^*$ bound state with $J^P={1}/{2}^{-}$, and the $|{}^2\mathbb{S}_{{1}/{2}}\rangle$ channel dominates with probability greater than 99\%. Furthermore, we consider the coupled channel effect arising from the $\Omega_{c} D_s^*$ and $\Omega_{c}^{*} D_s^*$ channels to discuss the bound state properties for the $\Omega_{c}{D}_s^*$ state with $J^P={1}/{2}^{-}$. By taking the cutoff value $\Lambda$ around 1.29 GeV, the bound state solutions can be obtained, where the $\Omega_{c} D_s^*$ channel has the highest probability with almost 86\% probability. Given that the $\Omega_{c}{D}_s^*$ bound state with $J^P={1}/{2}^{-}$ exist the small binding energy and the appropriate size within the reasonable cutoff value, we conclude that the $\Omega_{c}{D}_s^*$ state with $J^P={1}/{2}^{-}$ can be considered as the most promising doubly charmed molecular pentaquark candidate.

To obtain the bound state solutions for the $\Omega_{c}{D}_s^*$ state with $J^P={3}/{2}^{-}$, we set the cutoff value $\Lambda$ to be approximately 2.34 GeV during the single channel analysis. By taking into account the contribution of the $S$-$D$ wave mixing effect, we can obtain the bound state solutions with the cutoff value $\Lambda$ above 2.31 GeV, where the dominant channel is the $|{}^4\mathbb{S}_{{3}/{2}}\rangle$ with probability greater than 99\% and the $D$-wave channels have very small probabilities. Considering the contribution of the coupled channel effect and setting the cutoff value $\Lambda$ at approximately 1.50 GeV, we can obtain the bound state solutions, and the contribution of the coupled channel effect is significant in the formation of the $\Omega_{c}{D}_s^*$ bound state with $J^P={3}/{2}^{-}$, where the contribution from the $\Omega_{c}^{*} D_s^*$ channel is also obvious, apart from the $\Omega_{c} D_s^*$ channel. After comparing the obtained bound state properties of the $\Omega_{c}{D}_s^*$ state with $J^P={1}/{2}^{-}$, there is no priority for the $\Omega_{c}{D}_s^*$ state with $J^P={3}/{2}^{-}$ as the most promising doubly charmed molecular pentaquark candidate. However, the $\Omega_{c}{D}_s^*$ state with $J^P={3}/{2}^{-}$ can be regarded as the possible doubly charmed molecular pentaquark candidate.

\subsubsection{Bound state properties of the $\Omega_{c}^*{D}_s^*$ system}

The total effective potential of the $\Omega_{c}^*{D}_s^*$ system contains the exchange potentials of the $\eta$, $f_0$, and $\phi$ within the OBE model, while there exists the tensor force in the $\eta$ and $\phi$ exchange potentials, which results in the $S$-$D$ wave mixing effect contributing to the bound state properties of the $\Omega_{c}^{*} D_s^*$ system. The bound state properties of the $\Omega_{c}^{*} D_s^*$ system can be discussed by numerically solving the coupled channel Schr\"{o}dinger equation, and the corresponding bound state solutions of the $\Omega_{c}^{*} D_s^*$ system are listed in Table \ref{massspectra4}.

\renewcommand\tabcolsep{0.07cm}
\renewcommand{\arraystretch}{1.50}
\begin{table}[!htbp]
\caption{The obtained bound state solutions for the $\Omega_{c}^{*} D_s^*$ system. The units of the cutoff parameter $\Lambda$, the binding energy $E$, and the root-mean-square radius $r_{\rm RMS}$ are $ \rm{GeV}$, $\rm {MeV}$, and $\rm {fm}$, respectively. The primary probability of the corresponding channels is highlighted in bold font.}\label{massspectra4}
\begin{tabular}{c|ccc|cccc}\toprule[1pt]\toprule[1pt]
$J^P$&\multicolumn{3}{c|}{Single channel case}&\multicolumn{4}{c}{$S$-$D$ wave mixing case}\\\midrule[1.0pt]
\multirow{4}{*}{$\frac{1}{2}^{-}$}&$\Lambda$ &$E$  &$r_{\rm RMS}$ &$\Lambda$ &$E$  &$r_{\rm RMS}$ &P(${}^2\mathbb{S}_{\frac{1}{2}}/{}^4\mathbb{D}_{\frac{1}{2}}/{}^6\mathbb{D}_{\frac{1}{2}})$\\
\cline{2-8}
&$1.27$&$-0.37$&$4.27$&      1.27&$-0.82$ &3.15&\textbf{99.83}/0.10/0.07\\
                              &$1.29$&$-3.76$&$1.53$&      1.29&$-4.70$ &1.39&\textbf{99.73}/0.16/0.11\\
                              &$1.31$&$-10.39$&$0.96$&     1.31&$-11.61$ &0.92&\textbf{99.72}/0.17/0.12\\\hline
\multirow{4}{*}{$\frac{3}{2}^{-}$}&$\Lambda$ &$E$  &$r_{\rm RMS}$ &$\Lambda$ &$E$  &$r_{\rm RMS}$ &P(${}^4\mathbb{S}_{\frac{3}{2}}/{}^2\mathbb{D}_{\frac{3}{2}}/{}^4\mathbb{D}_{\frac{3}{2}}/{}^6\mathbb{D}_{\frac{3}{2}})$\\
\cline{2-8}
&1.46&$-0.53$ &3.81&      1.45&$-0.39$ &4.24&\textbf{99.87}/0.04/0.08/0.01\\
                              &1.50&$-5.15$ &1.36&      1.49&$-4.40$ &1.47&\textbf{99.74}/0.07/0.17/0.02\\
                              &1.53&$-11.70$ &0.94&     1.52&$-10.27$ &1.00&\textbf{99.73}/0.08/0.18/0.02\\\hline
\multirow{4}{*}{$\frac{5}{2}^{-}$}&$\Lambda$ &$E$  &$r_{\rm RMS}$ &$\Lambda$ &$E$  &$r_{\rm RMS}$ &P(${}^6\mathbb{S}_{\frac{5}{2}}/{}^2\mathbb{D}_{\frac{5}{2}}/{}^4\mathbb{D}_{\frac{5}{2}}/{}^6\mathbb{D}_{\frac{5}{2}})$\\
\cline{2-8}
&2.24&$-0.31$ &4.28&      2.22&$-0.29$ &4.86&\textbf{99.94}/$o(0)$/$o(0)$/0.06\\
                              &2.37&$-1.40$ &2.75&      2.36&$-1.50$ &2.67&\textbf{99.91}/0.01/$o(0)$/0.08\\
                              &2.50&$-3.20$ &1.94&     2.50&$-3.70$ &1.83&\textbf{99.89}/0.02/$o(0)$/0.09\\
\bottomrule[1pt]\bottomrule[1pt]
\end{tabular}
\end{table}

From the bound state properties for the $\Omega_{c}^{*} D_s^*$ system listed in Table \ref{massspectra4}, several noteworthy results can be inferred:
\begin{itemize}
  \item For the $\Omega_{c}^*{D}_s^*$ state with $J^P={1}/{2}^{-}$, the bound state solutions exist when the cutoff value $\Lambda$ is taken around 1.27 GeV if only considering the contribution of the $S$-wave component. In the analysis of the $S$-$D$ wave mixing, it was observed that the corresponding cutoff value reduces when the same binding energy is reached. Additionally, the total probability of the $D$-wave components for this bound state is less than 1\%. Based on the theoretical analysis of the bound state properties of the deuteron, the $\Omega_{c}^*{D}_s^*$ state with $J^P={1}/{2}^{-}$ can be viewed as the most promising doubly charmed molecular pentaquark candidate.
  \item For the $\Omega_{c}^*{D}_s^*$ state with $J^P={3}/{2}^{-}$, the bound state solutions with the cutoff value $\Lambda$ around 1.46 GeV can be obtained by conducting the single channel analysis. By taking into account the contribution from the $D$-wave channels, we can find the bound state solutions with the cutoff value $\Lambda=1.45~{\rm GeV}$, and the probability of the $|{}^4\mathbb{S}_{{3}/{2}}\rangle$ channel is over 99\%. Hence, we propose the $\Omega_{c}^*{D}_s^*$ state with $J^P={3}/{2}^{-}$ as the most promising doubly charmed molecular pentaquark candidate.
  \item For the $\Omega_{c}^*{D}_s^*$ state with $J^P={5}/{2}^{-}$, we can obtain the bound state solutions by performing the single channel analysis when the cutoff value $\Lambda$ is slightly larger than 2.24 GeV. In addition, the bound state solutions exist when the cutoff value $\Lambda$ is tuned larger than 2.22 GeV after considering the $S$-$D$ wave mixing effect, but the contribution of the $D$ wave channels is very small. Since the cutoff parameter $\Lambda$ for the formation of the $\Omega_{c}^*{D}_s^*$ bound state with $J^P={5}/{2}^{-}$ is obviously far from the reasonable range around 1.0 GeV, we cannot support the existence of the doubly charmed molecular pentaquark candidate for the $\Omega_{c}^*{D}_s^*$ state with $J^P={5}/{2}^{-}$.
\end{itemize}

Based on the obtained numerical results, we can predict the $\Omega_{c}{D}_s^*$ state with $J^P={1}/{2}^{-}$, the $\Omega_{c}^*{D}_s^*$ state with $J^P={1}/{2}^{-}$, and the $\Omega_{c}^*{D}_s^*$ state with $J^P={3}/{2}^{-}$ as the most promising doubly charmed molecular pentaquark candidates, since they exhibit the small binding energies and the suitable sizes within the reasonable cutoff values. In addition, the $\Omega_{c}{D}_s$ state with $J^P={1}/{2}^{-}$, the $\Omega_{c}^*{D}_s$ state with $J^P={3}/{2}^{-}$, and the $\Omega_{c}{D}_s^*$ state with $J^P={3}/{2}^{-}$ are the possible doubly charmed molecular pentaquark candidates. Compared to the $\Omega_{c}^{(*)}\bar {D}_s^{(*)}$ systems \cite{Wang:2021hql}, the $\Omega_{c}^{(*)} {D}_s^{(*)}$ systems have more hadronic molecular candidates, which is similar to the situation with the $\Xi_c^{(\prime,\,*)}\bar D_s^{(*)}$ systems \cite{Wang:2020bjt} and the $\Xi_c^{(\prime,\,*)}D_s^{(*)}$ systems \cite{Yalikun:2023waw}.

\section{The exploration of radiative decays and magnetic moments of the $\Omega_{c}^{(*)}{D}_s^*$ molecules}\label{sec3}

In the previous section, we already predicted the mass spectra of the $\Omega_{c}^{(*)}{D}_s^{(*)}$-type doubly charmed molecular pentaquarks, and our analysis concluded that the $\Omega_{c}{D}_s^*$ state with $J^P={1}/{2}^{-}$, the $\Omega_{c}^*{D}_s^*$ state with $J^P={1}/{2}^{-}$, and the $\Omega_{c}^*{D}_s^*$ state with $J^P={3}/{2}^{-}$ can be considered as the most promising doubly charmed molecular pentaquark candidates. However, differentiating the spin-parity quantum numbers for the $\Omega_{c}^*{D}_s^*$ molecules is the critical issue. This section discusses their radiative decays and magnetic moments, which may provide the crucial information to reveal their inner structures and aid in establishing the mass spectra of the $\Omega_{c}^{(*)}{D}_s^{(*)}$-type doubly charmed molecular pentaquarks. In the realistic calculations, the constituent quark model is used and the convention adopted is the same as in Refs. \cite{Wang:2022nqs,Wang:2023bek} to study the radiative decays and the magnetic moments of the $\Omega_{c}{D}_s^*$ molecule with $J^P={1}/{2}^{-}$, the $\Omega_{c}^*{D}_s^*$ molecule with $J^P={1}/{2}^{-}$, and the $\Omega_{c}^*{D}_s^*$ molecule with $J^P={3}/{2}^{-}$.

\subsection{The calculation method of radiative decays and magnetic moments of the hadrons within the constituent quark model}

As the important input information for studying the radiative decay widths between the hadronic states, it is essential to compute their transition magnetic moments \cite{Dey:1994qi,Simonis:2018rld,Gandhi:2019bju,Hazra:2021lpa,Li:2021ryu,Zhou:2022gra,Wang:2022tib,Rahmani:2020pol,Menapara:2022ksj,Menapara:2021dzi,Gandhi:2018lez,Majethiya:2011ry,Majethiya:2009vx,Shah:2016nxi,Ghalenovi:2018fxh,Wang:2022nqs,Mohan:2022sxm,An:2022qpt,Kakadiya:2022pin,Wang:2023bek}. In this work, we employ the constituent quark model to calculate the transition magnetic moments between the $\Omega_{c}{D}_s^*$ molecule with $J^P={1}/{2}^{-}$, the $\Omega_{c}^*{D}_s^*$ molecule with $J^P={1}/{2}^{-}$, and the $\Omega_{c}^*{D}_s^*$ molecule with $J^P={3}/{2}^{-}$.

Similar to the approach for dealing with the transition magnetic moments between the hadronic molecules discussed in Refs. \cite{Wang:2022nqs,Wang:2023bek}, the impact of the spatial wave functions of the initial and final states is also considered to investigate the transition magnetic moments between the $\Omega_{c}{D}_s^*$ molecule with $J^P={1}/{2}^{-}$, the $\Omega_{c}^*{D}_s^*$ molecule with $J^P={1}/{2}^{-}$, and the $\Omega_{c}^*{D}_s^*$ molecule with $J^P={3}/{2}^{-}$, which depend on their binding energies. Currently, there is a lack of the experimental information on the $\Omega_{c}^{(*)}{D}_s^{*}$ molecular states. To establish the transition magnetic moments between the $\Omega_{c}{D}_s^*$ molecule with $J^P={1}/{2}^{-}$, the $\Omega_{c}^*{D}_s^*$ molecule with $J^P={1}/{2}^{-}$, and the $\Omega_{c}^*{D}_s^*$ molecule with $J^P={3}/{2}^{-}$ in the present work, we assume the binding energies of the initial and final $\Omega_{c}^{(*)}{D}_s^{*}$ molecules to be identical for simplicity. In this context, the transition magnetic moments between the $\Omega_{c}{D}_s^*$ molecule with $J^P={1}/{2}^{-}$, the $\Omega_{c}^*{D}_s^*$ molecule with $J^P={1}/{2}^{-}$, and the $\Omega_{c}^*{D}_s^*$ molecule with $J^P={3}/{2}^{-}$ can be deduced by calculating the expectation values of the $z$-component of the magnetic moment operator, which can be expressed as \cite{Li:2021ryu,Zhou:2022gra,Wang:2022tib,Wang:2022nqs,Majethiya:2009vx,Majethiya:2011ry,Shah:2016nxi,Gandhi:2018lez,Simonis:2018rld,Ghalenovi:2018fxh,Gandhi:2019bju,Rahmani:2020pol,Hazra:2021lpa,Menapara:2021dzi,Menapara:2022ksj,Kakadiya:2022pin,Wang:2022nqs,Wang:2023bek}
\begin{eqnarray}
\mu_{H \to H^{\prime}}=\left\langle{J_{H^{\prime}},J_{z}\left|\sum_{j}\hat{\mu}_{zj}^{\rm spin}e^{-i {\bf k}\cdot{\bf r}_j}+\hat{\mu}_z^{\rm orbital}\right|J_{H},J_{z}}\right\rangle.
\end{eqnarray}
Here, $\mu_{H \to H^{\prime}}$ is the transition magnetic moment between the hadrons $H$ and $H^{\prime}$, $J_z$ is the lowest value between $J_H$ and $J_{H^{\prime}}$, the spatial wave function of the emitted photon for the $H \rightarrow H^{\prime}\gamma$ process is denoted by $e^{-i{\bf k} \cdot{\bf r}_j}$, and ${\bf k}$ refers to the momentum of the emitted photon, which is $k={(m_{H}^2-m_{H^{\prime}}^2)}/{2m_{H}}$. Within the constituent quark model, the magnetic moment operator is composed of the spin magnetic moment operator and the orbital magnetic moment operator, which can be written explicitly as \cite{Liu:2003ab,Huang:2004tn,Zhu:2004xa,Haghpayma:2006hu,Wang:2016dzu,Deng:2021gnb,Gao:2021hmv,Zhou:2022gra,Wang:2022tib,Li:2021ryu,Schlumpf:1992vq,Schlumpf:1993rm,Cheng:1997kr,Ha:1998gf,Ramalho:2009gk,Girdhar:2015gsa,Menapara:2022ksj,Mutuk:2021epz,Menapara:2021vug,Menapara:2021dzi,Gandhi:2018lez,Dahiya:2018ahb,Kaur:2016kan,Thakkar:2016sog,Shah:2016vmd,Dhir:2013nka,Sharma:2012jqz,Majethiya:2011ry,Sharma:2010vv,Dhir:2009ax,Simonis:2018rld,Ghalenovi:2014swa,Kumar:2005ei,Rahmani:2020pol,Hazra:2021lpa,Gandhi:2019bju,Majethiya:2009vx,Shah:2016nxi,Shah:2018bnr,Ghalenovi:2018fxh,Wang:2022nqs,Mohan:2022sxm,An:2022qpt,Kakadiya:2022pin,Wu:2022gie,Wang:2023bek}
\begin{eqnarray}
\hat{\mu}_{zj}^{\rm spin}&=&\frac{e_j}{2m_j}\hat{\sigma}_{zj},\\
\hat{\mu}_z^{\rm orbital}&=&\left(\frac{m_{m}}{m_{b}+m_{m}}\frac{e_b}{2m_b}+\frac{m_{b}}{m_{b}+m_{m}}\frac{e_m}{2m_m}\right)\hat{L}_z.
\end{eqnarray}
In the above expressions, we define the charge, the mass, and the $z$-component of the Pauli spin operator of the $j$-th constituent of the hadron as $e_j$, $m_j$, and $\hat{\sigma}_{zj}$, respectively. In addition, the subscripts $b$ and $m$ have been utilized to distinguish the baryon $\Omega_{c}^{(*)}$ and the meson ${D}_s^{*}$, while $\hat{L}_z$ is the $z$-component of the orbital angular momentum operator between the constituent
hadrons $\Omega_{c}^{(*)}$ and ${D}_s^{*}$. When only considering the $S$-wave component, there is no contribution of the orbital magnetic moment to the transition magnetic moments between the hadrons.

On the basis of the transition magnetic moments among the $\Omega_{c}{D}_s^*$ molecule with $J^P={1}/{2}^{-}$, the $\Omega_{c}^*{D}_s^*$ molecule with $J^P={1}/{2}^{-}$, and the $\Omega_{c}^*{D}_s^*$ molecule with $J^P={3}/{2}^{-}$, we can further discuss their radiative decay widths. The radiative decay width $\Gamma_{H \to H^{\prime}\gamma}$ can be linked to the transition magnetic moment $\mu_{H \to H^{\prime}}$, the momentum of the emitted photon $k$, the proton mass $m_p$, the electromagnetic fine structure constant $\alpha_{\rm {EM}}$, as well as the quantum numbers of the initial and final hadrons \cite{Dey:1994qi,Simonis:2018rld,Gandhi:2019bju,Hazra:2021lpa,Li:2021ryu,Zhou:2022gra,Wang:2022tib,Rahmani:2020pol,Menapara:2022ksj,Menapara:2021dzi,Gandhi:2018lez,Majethiya:2011ry,Majethiya:2009vx,Shah:2016nxi,Ghalenovi:2018fxh,Wang:2022nqs,Mohan:2022sxm,An:2022qpt,Kakadiya:2022pin,Wang:2023bek}, where the specific expression can be provided as follows \cite{Wang:2022nqs,Wang:2023bek}
\begin{eqnarray}
 \Gamma_{H \to H^{\prime}\gamma}=\frac{k^{3}}{m_{p}^{2}} \frac{\alpha_{\rm {EM}}}{2J_{H}+1}\frac{\sum\limits_{J_{H^{\prime}z},J_{Hz}}\left(\begin{array}{ccc} J_{H^{\prime}}&1&J_{H}\\-J_{H^{\prime}z}&0&J_{Hz}\end{array}\right)^2}{\left(\begin{array}{ccc} J_{H^{\prime}}&1&J_{H}\\-J_{z}&0&J_{z}\end{array}\right)^2}\frac{\left|\mu_{H \to H^{\prime}}\right|^2}{\mu_N^2}.\nonumber\\
\end{eqnarray}
Here, $m_p=0.938\,\mathrm{GeV}$ \cite{ParticleDataGroup:2022pth}, $\alpha_{\rm {EM}} \approx {1}/{137}$, $\mu_N=e/2m_p$, and the constants $\left(\begin{array}{ccc} J_{H^{\prime}}&1&J_{H}\\-J_{H^{\prime}z}&0&J_{Hz}\end{array}\right)$ and $\left(\begin{array}{ccc} J_{H^{\prime}}&1&J_{H}\\-J_{z}&0&J_{z}\end{array}\right)$ are the 3-$j$ coefficients.

In the present work, we also explore the magnetic moments of the $\Omega_{c}{D}_s^*$ molecule with $J^P={1}/{2}^{-}$, the $\Omega_{c}^*{D}_s^*$ molecule with $J^P={1}/{2}^{-}$, and the $\Omega_{c}^*{D}_s^*$ molecule with $J^P={3}/{2}^{-}$. Similar to the method used for estimating the transition magnetic moments between the hadrons, the magnetic moments of the hadrons $\mu_{H}$ within the constituent quark model can be calculated using the following expectation values \cite{Liu:2003ab,Huang:2004tn,Zhu:2004xa,Haghpayma:2006hu,Wang:2016dzu,Deng:2021gnb,Gao:2021hmv,Zhou:2022gra,Wang:2022tib,Li:2021ryu,Schlumpf:1992vq,Schlumpf:1993rm,Cheng:1997kr,Ha:1998gf,Ramalho:2009gk,Girdhar:2015gsa,
Menapara:2022ksj,Mutuk:2021epz,Menapara:2021vug,Menapara:2021dzi,Gandhi:2018lez,Dahiya:2018ahb,Kaur:2016kan,Thakkar:2016sog,Shah:2016vmd,Dhir:2013nka,Sharma:2012jqz,Majethiya:2011ry,Sharma:2010vv,Dhir:2009ax,Simonis:2018rld,
Ghalenovi:2014swa,Kumar:2005ei,Rahmani:2020pol,Hazra:2021lpa,Gandhi:2019bju,Majethiya:2009vx,Shah:2016nxi,Shah:2018bnr,Ghalenovi:2018fxh,Wang:2022nqs,Mohan:2022sxm,An:2022qpt,Kakadiya:2022pin,Wu:2022gie,Wang:2023bek}
\begin{eqnarray}
\mu_{H}&=&\left\langle{J_{H},J_{H}\left|\sum_{j}\hat{\mu}_{zj}^{\rm spin}+\hat{\mu}_z^{\rm orbital}\right|J_{H},J_{H}}\right\rangle.
\end{eqnarray}

\subsection{The results of radiative decays of the $\Omega_{c}^{(*)}{D}_s^*$ molecules}

The radiative decay width is a significant electromagnetic property of the hadron, which can be experimentally studied. Thus, we first study the radiative decays between the $\Omega_{c}{D}_s^*$ molecule with $J^P={1}/{2}^{-}$, the $\Omega_{c}^*{D}_s^*$ molecule with $J^P={1}/{2}^{-}$, and the $\Omega_{c}^*{D}_s^*$ molecule with $J^P={3}/{2}^{-}$. As the essential inputs for studying the radiative decay widths between the $\Omega_{c}{D}_s^*$ molecule with $J^P={1}/{2}^{-}$, the $\Omega_{c}^*{D}_s^*$ molecule with $J^P={1}/{2}^{-}$, and the $\Omega_{c}^*{D}_s^*$ molecule with $J^P={3}/{2}^{-}$, it is necessary to calculate their transition magnetic moments.

As demonstrated in Refs. \cite{Wang:2022tib,Zhou:2022gra,Li:2021ryu,Wang:2022nqs,Wang:2023bek}, the transition magnetic moments of the hadronic molecular states are composed of the linear combination of the magnetic moments and the transition magnetic moments of their constituent hadrons in the constituent quark model. Currently, the values of the magnetic moments and the transition magnetic moment of the hadrons $\Omega_{c}$, $\Omega_{c}^{*}$, and ${D}_s^{*}$ have not been determined experimentally. In the following, we discuss the magnetic moments and the transition magnetic moment of the hadrons $\Omega_{c}$, $\Omega_{c}^{*}$, and ${D}_s^{*}$. In order to calculate the magnetic moments and the transition magnetic moment of the hadrons $\Omega_{c}$, $\Omega_{c}^{*}$, and ${D}_s^{*}$ using the constituent quark model, it is necessary to construct their wave functions. The flavor wave functions of the $\Omega_{c}$, $\Omega_{c}^{*}$, and ${D}_s^{*}$ have the form of $ssc$, $ssc$, and $c \bar s$, respectively. In Table~\ref{spinwavefunctions}, we present the spin wave functions $|S,S_3\rangle$ of the $\Omega_{c}$, $\Omega_{c}^{*}$, and ${D}_s^{*}$, where $S$ and $S_3$ in the spin wave functions represent the total spins and the total spin third components of the $\Omega_{c}$, $\Omega_{c}^{*}$, and ${D}_s^{*}$, respectively.

\renewcommand\tabcolsep{0.49cm}
\renewcommand{\arraystretch}{1.50}
\begin{table}[!htbp]
\caption{The spin wave functions $|S,S_3\rangle$ of the $\Omega_{c}$, $\Omega_{c}^{*}$, and ${D}_s^{*}$. Here, $\uparrow$ and $\downarrow$ in the spin wave functions represent ${1}/{2}$ and $-{1}/{2}$ for the third components of the quark spins, respectively.}\label{spinwavefunctions}
\begin{tabular}{c|ll}\toprule[1pt]\toprule[1pt]
Hadrons&$|S,S_3\rangle$&The spin wave functions\\\midrule[1.0pt]
\multirow{2}{*}{$\Omega_{c}$}&$\left|{1}/{2},{1}/{2}\right\rangle$&$\frac{1}{\sqrt{6}}\left(2\uparrow\uparrow\downarrow-\downarrow\uparrow\uparrow-\uparrow\downarrow\uparrow\right)$\\
    &$\left|{1}/{2},-{1}/{2}\right\rangle$&$\frac{1}{\sqrt{6}}\left(\downarrow\uparrow\downarrow+\uparrow\downarrow\downarrow-2\downarrow\downarrow\uparrow\right)$\\\hline
\multirow{4}{*}{$\Omega_{c}^{*}$}& $\left|{3}/{2},{3}/{2}\right\rangle$&$\uparrow\uparrow\uparrow$\\
    &$\left|{3}/{2},{1}/{2}\right\rangle$&$\frac{1}{\sqrt{3}}\left(\downarrow\uparrow\uparrow+\uparrow\downarrow\uparrow+\uparrow\uparrow\downarrow\right)$\\
     &$\left|{3}/{2},-{1}/{2}\right\rangle$&$\frac{1}{\sqrt{3}}\left(\downarrow\downarrow\uparrow+\uparrow\downarrow\downarrow+\downarrow\uparrow\downarrow\right)$\\
    &$\left|{3}/{2},-{3}/{2}\right\rangle$&$\downarrow\downarrow\downarrow$\\\hline
\multirow{3}{*}{$D^{*}_s$}& $\left|1,1\right\rangle$&$\uparrow\uparrow$\\
    &$\left|1,0\right\rangle$&$\frac{1}{\sqrt{2}}\left(\uparrow\downarrow+\downarrow\uparrow\right)$\\
    &$\left|1,-1\right\rangle$&$\downarrow\downarrow$\\
\bottomrule[1pt]\bottomrule[1pt]
\end{tabular}
\end{table}

In addition to their flavor and spin wave functions, we also need to discuss the spatial wave functions of the $\Omega_{c}$ and $\Omega_{c}^{*}$ baryons when calculating the transition magnetic moment between the hadrons $\Omega_{c}$ and $\Omega_{c}^{*}$. In our current study, we utilize the simple harmonic oscillator wave function $\phi_{n,l,m}(\beta,{\bf r})$ to describe the spatial wave functions of the baryons $\Omega_{c}$ and $\Omega_{c}^{*}$ \cite{Wang:2022nqs,Wang:2023bek}, and the simple harmonic oscillator wave function can be defined as follows
\begin{eqnarray}
\phi_{n,l,m}(\beta,{\bf r})&=&\sqrt{\frac{2n!}{\Gamma(n+l+\frac{3}{2})}}L_{n}^{l+\frac{1}{2}}(\beta^2r^2)\beta^{l+\frac{3}{2}}{\mathrm e}^{-\frac{\beta^2r^2}{2}}r^l Y_{l m}(\Omega_{\bf r}).\nonumber\\
\end{eqnarray}
Here, we define the radial, the orbital, and the magnetic quantum numbers of the discussed hadron as $n$, $l$, and $m$, respectively. Additionally, $L_{n}^{l+\frac{1}{2}}(x)$ denotes the associated Laguerre polynomial, $Y_{l m}(\Omega_{\bf r})$ represents the spherical harmonic function, and $\beta$ is a parameter in the simple harmonic oscillator wave function. Our previous theoretical study \cite{Wang:2022nqs} utilized $\beta = 0.4~{\rm GeV}$ to investigate the transition magnetic moments of the $\Omega_c^{(*)} \bar D_s^*$ molecules, and this value is also utilized in the present work. In practice, when calculating the contribution of the spatial wave functions of the initial and final states $\left\langle \phi_f \left|e^{-i {\bf k}\cdot{\bf r}_j}\right| \phi_i\right\rangle$, the spatial wave function of the emitted photon $e^{-i{\bf k}\cdot{\bf r}_j}$ needs to be expanded using the spherical Bessel function $j_l(x)$ and the spherical harmonic function $Y_{l m}(\Omega_{\bf r})$ \cite{Khersonskii:1988krb}
\begin{eqnarray}
e^{-i{\bf k}\cdot{\bf r}_j}&=&\sum\limits_{l=0}^\infty\sum\limits_{m=-l}^l4\pi(-i)^lj_l(kr_j)Y_{lm}^*(\Omega_{\bf k})Y_{lm}(\Omega_{{\bf r}_j}).
\end{eqnarray}
Utilizing the aforementioned discussions, the contribution of the spatial wave functions of the initial and final states $\left\langle \phi_f \left|e^{-i {\bf k}\cdot{\bf r}_j}\right| \phi_i\right\rangle$ can be calculated \cite{Wang:2022nqs}.

Taking into account the flavor, spin, and spatial wave functions of the $\Omega_c^{(*)}$ baryons and the $D^{*}_s$ meson, the magnetic moments and the transition magnetic moment of the $\Omega_c^{(*)}$ baryons and the $D^{*}_s$ meson can be deduced by calculating the expectation values of the $z$-component of the magnetic moment operator. Regarding the magnetic moments of the $\Omega_c^{(*)}$ baryons and the $D^{*}_s$ meson, we find $\mu_{\Omega^{0}_c}=\frac{4}{3}\mu_s-\frac{1}{3}\mu_c$, $\mu_{\Omega^{*0}_c}=2\mu_s+\mu_c$, and $\mu_{D^{*+}_s}=\mu_{c}+\mu_{\bar s}$. Moreover, we acquire $\mu_{\Omega^{*0}_c \to \Omega^{0}_c}=\frac{2\sqrt{2}}{3}\mu_s-\frac{2\sqrt{2}}{3}\mu_c$ if we disregard the contribution of the spatial wave functions of the initial and final states. Here, we take $\mu_{q}=-\mu_{\bar q}={e_q}/{2M_q}$. To provide quantitative description of the magnetic moments and the transition magnetic moment of $\Omega_c^{(*)}$ baryons and $D^{*}_s$ mesons, we take the constituent quark masses as $m_{s}=0.45\,\mathrm{GeV}$ and $m_{c}=1.68\,\mathrm{GeV}$ employed in prior work \cite{Kumar:2005ei}, which have been widely used in the last decades to investigate the hadronic magnetic moments and transition magnetic moments \cite{Li:2021ryu,Zhou:2022gra,Wang:2022tib,Wang:2022nqs}. In Table~\ref{MT2}, we list the obtained numerical results for the magnetic moments and the transition magnetic moment of the $\Omega_c^{(*)}$ baryons and the $D^{*}_s$ meson.

\renewcommand\tabcolsep{0.03cm}
\renewcommand{\arraystretch}{1.50}
\begin{table}[!htbp]
  \caption{The numerical results for the magnetic moments and the transition magnetic moment of the $\Omega_c^{(*)}$ baryons and the $D^{*}_s$ meson. The units of the magnetic moments and the transition magnetic moments are $\mu_N=e/2m_p$. For the $\mu_{\Omega^{*0}_c \to \Omega^{0}_c}$, we consider the contribution of the initial and final state spatial wave functions.}\label{MT2}
\begin{tabular}{c|c|l}
\toprule[1.0pt]
\toprule[1.0pt]
Electromagnetic properties &  \multicolumn{1}{c|}{Present work}  &  \multicolumn{1}{c}{Other works} \\\midrule[1.0pt]
$\mu_{\Omega^{0}_c}$                   & $-1.051$        & $-1.127$ \cite{Gandhi:2018lez},\,$-0.960$ \cite{Patel:2007gx} \\
$\mu_{\Omega^{*0}_c}$                   & $-1.018$                            & $-1.127$ \cite{Gandhi:2018lez},\,$-0.936$ \cite{Simonis:2018rld}\\
$\mu_{D^{*+}_s}$                           & $1.067$                            &$1.000$ \cite{Simonis:2018rld},\,$1.080$ \cite{Zhang:2021yul}\\
$\mu_{\Omega^{*0}_c \to \Omega^{0}_c}$    & $-1.003$      &$-0.960$ \cite{Sharma:2010vv},\,$-1.128$ \cite{Majethiya:2009vx}\\
\bottomrule[1.0pt]
\bottomrule[1.0pt]
\end{tabular}
\end{table}

In Table~\ref{MT2}, we also list the magnetic moments and the transition magnetic moment of the $\Omega_c^{(*)}$ baryons and the $D^{*}_s$ meson predicted in other theoretical works. Especially, our obtained magnetic moments and transition magnetic moment of the $\Omega_c^{(*)}$ baryons and the $D^{*}_s$ meson are consistent with those predicted in other theoretical works. Thus, we can get the reliable results of the transition magnetic moments between the $\Omega_{c}{D}_s^*$ molecule with $J^P={1}/{2}^{-}$, the $\Omega_{c}^*{D}_s^*$ molecule with $J^P={1}/{2}^{-}$, and the $\Omega_{c}^*{D}_s^*$ molecule with $J^P={3}/{2}^{-}$ by utilizing our obtained magnetic moments and transition magnetic moment of the $\Omega_c^{(*)}$ baryons and the $D^{*}_s$ meson.

In addition, we discuss the impact of the spatial wave functions of the initial and final states to the transition magnetic moment of the $\Omega^{*0}_c \to \Omega^{0}_c \gamma$ process. By disregarding the contribution of the spatial wave functions of the initial and final states, we reach a value of $\mu_{\Omega^{*0}_c \to \Omega^{0}_c}=-1.006\mu_N$, which is comparable to the result of $\mu_{\Omega^{*0}_c \to \Omega^{0}_c}=-1.003\mu_N$ when including the contribution of the spatial wave functions of the initial and final states. Here, we
need to mention that the momentum of the emitted photon is about $70~{\rm MeV}$ for the $\Omega^{*0}_c \to \Omega^{0}_c \gamma$ process, and the obtained factors $\left\langle \phi_f \left|e^{-i {\bf k}\cdot{\bf r}_j}\right| \phi_i\right\rangle$ are approximately 0.99. Furthermore, if the value of $\beta$ in the simple harmonic oscillator wave function grows from $0.3$ to $0. 5~{\rm GeV}$, the contribution of the spatial wave functions of the initial and final states $\left\langle \phi_f \left|e^{-i {\bf k}\cdot{\bf r}_j}\right| \phi_i\right\rangle$ for the $\Omega^{*0}_c \to \Omega^{0}_c \gamma$ process exhibit slight changes, which does not obviously affect the transition magnetic moment of the $\Omega^{*0}_c \to \Omega^{0}_c \gamma$ process.

In the subsequent analysis, we study the transition magnetic moments and the radiative decay widths between the $\Omega_{c}{D}_s^*$ molecule with $J^P={1}/{2}^{-}$, the $\Omega_{c}^*{D}_s^*$ molecule with $J^P={1}/{2}^{-}$, and the $\Omega_{c}^*{D}_s^*$ molecule with $J^P={3}/{2}^{-}$ using the single channel analysis, the $S$-$D$ wave mixing analysis, and the coupled channel analysis, respectively. From Table \ref{massspectra3}, it should be noted that the mass of the $\Omega_{c}^*{D}_s^*$ molecule with $J^P={3}/{2}^{-}$ is greater than that of the $\Omega_{c}^*{D}_s^*$ molecular state with $J^P={1}/{2}^{-}$ when utilizing the same cutoff parameter. Consequently, this research focuses on investigating the transition magnetic moments and the radiative decay widths of the $\Omega_c^* D_s^*|{1}/{2}^-\rangle \to \Omega_c D_s^*|{1}/{2}^-\rangle\gamma$, $\Omega_c^* D_s^*|{3}/{2}^-\rangle \to \Omega_c D_s^*|{1}/{2}^-\rangle\gamma$, and $\Omega_c^* D_s^*|{3}/{2}^-\rangle \to \Omega_c^* D_s^*|{1}/{2}^-\rangle\gamma$ processes.

For the $\Omega_{c}{D}_s^*$ state with $J^P={1}/{2}^{-}$, the $\Omega_{c}^*{D}_s^*$ state with $J^P={1}/{2}^{-}$, and the $\Omega_{c}^*{D}_s^*$ state with $J^P={3}/{2}^{-}$, their flavor wave functions $|I,I_3\rangle$ are $|0,0\rangle=|\Omega_{c}^{(*)0}{D}_s^{*+}\rangle$. In addition, their spin wave functions $|S,S_3\rangle$ can be constructed by coupling the spin wave functions of the corresponding constituent hadrons $\Omega_{c}^{(*)}$ and ${D}_s^{*}$. When calculating the transition magnetic moment of the $D$-wave channel, it is necessary to expand the corresponding spin-orbital wave function $|{ }^{2 S+1} L_{J}\rangle$ employing the spin wave function $\left|S, m_{S}\right\rangle$ and the orbital wave function $Y_{L m_{L}}$ \cite{Wang:2022tib,Zhou:2022gra,Li:2021ryu,Wang:2022nqs,Wang:2023bek}, i.e., $\left|{ }^{2 S+1} L_{J}\right\rangle=\sum_{m_{S}, m_{L}} C_{S m_{S}, L m_{L}}^{J,M} \left|S, m_{S}\right\rangle Y_{L m_{L}}$.
And then, the transition magnetic moment of the $D$-wave channel can be inferred by calculating the expectation value of the spin and orbital magnetic moment operators.

As mentioned above, the transition magnetic moments between the $\Omega_{c}{D}_s^*$ molecule with $J^P={1}/{2}^{-}$, the $\Omega_{c}^*{D}_s^*$ molecule with $J^P={1}/{2}^{-}$, and the $\Omega_{c}^*{D}_s^*$ molecule with $J^P={3}/{2}^{-}$ are dependent on the spatial wave functions of the initial and final $\Omega_{c}^{(*)}{D}_s^{*}$ molecules. In the present work, we take the precise spatial wave functions for the $\Omega_{c}^{(*)}{D}_s^{*}$ molecular states to discuss their transition magnetic moments, which can be obtained by studying their mass spectra in the above section. Currently, there is no available experimental information on the $\Omega_{c}^{(*)}{D}_s^{*}$ molecular states. However, it is noteworthy that their spatial wave functions are dependent on their binding energies. For the purpose of this investigation, we assume the same binding energies for the initial and final $\Omega_{c}^{(*)}{D}_s^{*}$ molecules and three binding energies $-0.5$, $-6.0$, and $-12.0$ MeV to discuss the transition magnetic moments between the $\Omega_{c}{D}_s^*$ molecule with $J^P={1}/{2}^{-}$, the $\Omega_{c}^*{D}_s^*$ molecule with $J^P={1}/{2}^{-}$, and the $\Omega_{c}^*{D}_s^*$ molecule with $J^P={3}/{2}^{-}$. Based on the obtained transition magnetic moments between the $\Omega_{c}{D}_s^*$ molecule with $J^P={1}/{2}^{-}$, the $\Omega_{c}^*{D}_s^*$ molecule with $J^P={1}/{2}^{-}$, and the $\Omega_{c}^*{D}_s^*$ molecule with $J^P={3}/{2}^{-}$, we can further calculate their radiative decay widths. In Table~\ref{ME4}, we collect the obtained numerical results of the transition magnetic moments and the radiative decay widths between the $\Omega_{c}{D}_s^*$ molecule with $J^P={1}/{2}^{-}$, the $\Omega_{c}^*{D}_s^*$ molecule with $J^P={1}/{2}^{-}$, and the $\Omega_{c}^*{D}_s^*$ molecule with $J^P={3}/{2}^{-}$ based on the single channel analysis, the $S$-$D$ wave mixing analysis, and the coupled channel analysis, respectively.

\renewcommand\tabcolsep{0.18cm}
\renewcommand{\arraystretch}{1.50}
\begin{table*}[!htbp]
  \caption{The obtained transition magnetic moments and radiative decay widths between the $\Omega_{c}{D}_s^*$ molecule with $J^P={1}/{2}^{-}$, the $\Omega_{c}^*{D}_s^*$ molecule with $J^P={1}/{2}^{-}$, and the $\Omega_{c}^*{D}_s^*$ molecule with $J^P={3}/{2}^{-}$. The units of the transition magnetic moments and the radiative decay widths between the hadrons are $\mu_N$ and ${\rm keV}$, respectively. Here, Case I, Case II, and Case III correspond to the results obtained based on the single channel analysis, the $S$-$D$ wave mixing analysis, and the coupled channel analysis, respectively.}\label{ME4}
\begin{tabular}{c|c|ccc}
\toprule[1.0pt]
\toprule[1.0pt]
Electromagnetic properties&Decay processes &  Case I& Case II & Case III\\\midrule[1.0pt]
\multirow{3}{*}{$\mu_{H \to H^{\prime}}$}&$\Omega_c^* D_s^*|{1}/{2}^-\rangle \to \Omega_c D_s^*|{1}/{2}^-\rangle\gamma$  & $0.629$,\,$0.664$,\,$0.665$& $0.629$,\,$0.663$,\,$0.665$&$0.538$,\,$0.369$,\,$0.265$\\
&$\Omega_c^* D_s^*|{3}/{2}^-\rangle \to \Omega_c D_s^*|{1}/{2}^-\rangle\gamma$  & $-0.702$,\,$-0.742$,\,$-0.744$& $-0.702$,\,$-0.741$,\,$-0.743$&$-0.804$,\,$-1.053$,\,$-1.152$\\
&$\Omega_c^* D_s^*|{3}/{2}^-\rangle \to \Omega_c^* D_s^*|{1}/{2}^-\rangle\gamma$  & $-1.301$,\,$-1.301$,\,$-1.301$& $-1.301$,\,$-1.300$,\,$-1.300$&/\\\hline
\multirow{2}{*}{$\Gamma_{H \to H^{\prime}\gamma}$}&$\Omega_c^* D_s^*|{1}/{2}^-\rangle \to \Omega_c D_s^*|{1}/{2}^-\rangle\gamma$  & $1.135$,\,$1.263$,\,$1.270$& $1.134$,\,$1.260$,\,$1.267$&$0.831$,\,$0.390$,\,$0.201$\\
&$\Omega_c^* D_s^*|{3}/{2}^-\rangle \to \Omega_c D_s^*|{1}/{2}^-\rangle\gamma$  & $0.708$,\,$0.789$,\,$0.793$& $0.706$,\,$0.787$,\,$0.791$&$0.926$,\,$1.590$,\,$1.902$\\
\bottomrule[1.0pt]
\bottomrule[1.0pt]
\end{tabular}
\end{table*}

According to the numerical results of the transition magnetic moments and the radiative decay widths between the $\Omega_{c}{D}_s^*$ molecule with $J^P={1}/{2}^{-}$, the $\Omega_{c}^*{D}_s^*$ molecule with $J^P={1}/{2}^{-}$, and the $\Omega_{c}^*{D}_s^*$ molecule with $J^P={3}/{2}^{-}$ listed in Table~\ref{ME4}, we want to specify four points: (i) The $S$-$D$ wave mixing effect plays a minor role in modifying the decay widths of the $\Omega_c^* D_s^*|{1}/{2}^-\rangle \to \Omega_c D_s^*|{1}/{2}^-\rangle\gamma$ and $\Omega_c^* D_s^*|{3}/{2}^-\rangle \to \Omega_c D_s^*|{1}/{2}^-\rangle\gamma$ processes, but the contribution of the coupled channel effect is crucial in mediating the decay widths of the $\Omega_c^* D_s^*|{1}/{2}^-\rangle \to \Omega_c D_s^*|{1}/{2}^-\rangle\gamma$ and $\Omega_c^* D_s^*|{3}/{2}^-\rangle \to \Omega_c D_s^*|{1}/{2}^-\rangle\gamma$ processes. (ii) The decay widths of the $\Omega_c^* D_s^*|{1}/{2}^-\rangle \to \Omega_c D_s^*|{1}/{2}^-\rangle\gamma$ and $\Omega_c^* D_s^*|{3}/{2}^-\rangle \to \Omega_c D_s^*|{1}/{2}^-\rangle\gamma$ processes are influenced by the binding energies of the initial and final $\Omega_{c}^{(*)}{D}_s^{*}$ molecules, particularly in the case of the coupled channel analysis. Thus, we strongly expect that the binding energies of the $\Omega_{c}{D}_s^*$ molecule with $J^P={1}/{2}^{-}$, the $\Omega_{c}^*{D}_s^*$ molecule with $J^P={1}/{2}^{-}$, and the $\Omega_{c}^*{D}_s^*$ molecule with $J^P={3}/{2}^{-}$ can be established in the forthcoming experiments, which can provide the important inputs to improve our understanding of the decay widths of the $\Omega_c^* D_s^*|{1}/{2}^- \rangle \to \Omega_c D_s^*|{1}/{2}^-\rangle\gamma$ and $\Omega_c^* D_s^*|{3}/{2}^-\rangle \to \Omega_c D_s^*|{1}/{2}^-\rangle\gamma$ processes. (iii) Combined with the information on the binding energies of the $\Omega_{c}^*{D}_s^{(*)}$ molecular states and the radiative decay behaviors of the $\Omega_c^* D_s^*|{1}/{2}^-\rangle \to \Omega_c D_s^*|{1}/{2}^-\rangle\gamma$ and $\Omega_c^* D_s^*|{3}/{2}^-\rangle \to \Omega_c D_s^*|{1}/{2}^-\rangle\gamma$ processes, the spin-parity quantum numbers of the $\Omega_{c}^*{D}_s^*$ molecular states can be distinguished. In the future, we urge our theoretical and experimental colleagues to study the radiative decay behaviors of the $\Omega_c^* D_s^*|{1}/{2}^-\rangle \to \Omega_c D_s^*|{1}/{2}^-\rangle\gamma$ and $\Omega_c^* D_s^*|{3}/{2}^-\rangle \to \Omega_c D_s^*|{1}/{2}^-\rangle\gamma$ processes. (iv) Assuming the same binding energies for the $\Omega_{c}^*{D}_s^*$ molecule with $J^P={1}/{2}^{-}$ and the $\Omega_{c}^*{D}_s^*$ molecular state with $J^P={3}/{2}^{-}$, the phase space is zero for the $\Omega_c^* D_s^*|{3}/{2}^-\rangle \to \Omega_c^* D_s^*|{1}/{2}^-\rangle\gamma$ process, and we get $\Gamma_{\Omega_c^* D_s^*|{3}/{2}^-\rangle \to \Omega_c^* D_s^*|{1}/{2}^-\rangle\gamma}=0$. Taking different binding energies for the $\Omega_{c}^*{D}_s^*$ molecular state with $J^P={1}/{2}^{-}$ and the $\Omega_{c}^*{D}_s^*$ molecule with $J^P={3}/{2}^{-}$ and scanning their binding energies in the range from $-12$ to $-0.5$ MeV, we can estimate that the decay width of the $\Omega_c^* D_s^*|{3}/{2}^-\rangle \to \Omega_c^* D_s^*|{1}/{2}^-\rangle\gamma$ process is less than $0.011{\rm~keV}$. Specifically, the decay width of the $\Omega_c^* D_s^*|{3}/{2}^-\rangle \to \Omega_c^* D_s^*|{1}/{2}^-\rangle\gamma$ process is highly suppressed, and this is primarily due to the limited phase space available for this radiative decay process.

\subsection{The results of magnetic moments of the $\Omega_{c}^{(*)}{D}_s^*$ molecules}

In the previous subsection, we discussed the radiative decays of the $\Omega_{c}^{(*)}{D}_s^{*}$ molecules, which can provide the crucial information to reflect their inner structures and disentangle the spin-parity quantum numbers of the $\Omega_{c}^*{D}_s^*$ molecular states. In the following, we explore the magnetic moments of the $\Omega_{c}{D}_s^*$ molecule with $J^P={1}/{2}^{-}$, the $\Omega_{c}^*{D}_s^*$ molecule with $J^P={1}/{2}^{-}$, and the $\Omega_{c}^*{D}_s^*$ molecule with $J^P={3}/{2}^{-}$ using the single channel analysis, the $S$-$D$ wave mixing analysis, and the coupled channel analysis, respectively.

On the basis of the constructed flavor and spin wave functions of the $\Omega_{c}{D}_s^*$ state with $J^P={1}/{2}^{-}$, the $\Omega_{c}^*{D}_s^*$ state with $J^P={1}/{2}^{-}$, and the $\Omega_{c}^*{D}_s^*$ state with $J^P={3}/{2}^{-}$, we can deduce their magnetic moments by computing the expectation values of the $z$-component of the magnetic moment operator. Including the contribution of the $S$-$D$ wave mixing effect and the coupled channel effect, their magnetic moments depend on the spatial wave functions of the corresponding mixing channels. In the present work, we consider three  binding energies $-0.5$, $-6.0$, and $-12.0$ MeV for the $\Omega_{c}{D}_s^*$ molecule with $J^P={1}/{2}^{-}$, the $\Omega_{c}^*{D}_s^*$ molecule with $J^P={1}/{2}^{-}$, and the $\Omega_{c}^*{D}_s^*$ molecule with $J^P={3}/{2}^{-}$ to discuss the spatial wave functions of the relevant mixing channels, from which we can obtain their magnetic moments by considering the $S$-$D$ wave mixing effect and the coupled channel effect. In Table~\ref{Magneticmoments}, we give the numerical results of the magnetic moments of the $\Omega_{c}{D}_s^*$ molecule with $J^P={1}/{2}^{-}$, the $\Omega_{c}^*{D}_s^*$ molecule with $J^P={1}/{2}^{-}$, and the $\Omega_{c}^*{D}_s^*$ molecule with $J^P={3}/{2}^{-}$ obtained by the single channel analysis, the $S$-$D$ wave mixing analysis, and the coupled channel analysis, respectively.

\renewcommand\tabcolsep{0.10cm}
\renewcommand{\arraystretch}{1.50}
\begin{table}[!htbp]
  \caption{The obtained magnetic moments of the $\Omega_{c}{D}_s^*$ molecule with $J^P={1}/{2}^{-}$, the $\Omega_{c}^*{D}_s^*$ molecule with $J^P={1}/{2}^{-}$, and the $\Omega_{c}^*{D}_s^*$ molecule with $J^P={3}/{2}^{-}$. The units of the magnetic moments of the hadrons are $\mu_N$. Here, Case I, Case II, and Case III correspond to the results obtained based on the single channel analysis, the $S$-$D$ wave mixing analysis, and the coupled channel analysis, respectively.}
  \label{Magneticmoments}
\begin{tabular}{c|c|c|c}
\toprule[1.0pt]
\toprule[1.0pt]
Molecules &  Case I& Case II & Case III\\\midrule[1.0pt]
$\Omega_c D_s^*|{1}/{2}^-\rangle$  & $1.062$ & $1.061$,\,$1.061$,\,$1.061$& $1.124$,\,$1.186$,\,$1.169$\\
$\Omega_c^* D_s^*|{1}/{2}^-\rangle$  & $-0.921$ & $-0.921$,\,$-0.920$,\,$-0.920$&/\\
$\Omega_c^* D_s^*|{3}/{2}^-\rangle$  & $-0.319$ & $-0.319$,\,$-0.319$,\,$-0.319$&/\\
\bottomrule[1.0pt]
\bottomrule[1.0pt]
\end{tabular}
\end{table}

As shown in Table~\ref{Magneticmoments}, the magnetic moments of the $\Omega_{c}{D}_s^*$ molecule with $J^P={1}/{2}^{-}$, the $\Omega_{c}^*{D}_s^*$ molecule with $J^P={1}/{2}^{-}$, and the $\Omega_{c}^*{D}_s^*$ molecule with $J^P={3}/{2}^{-}$ are not affected by their binding energies in the single channel analysis, since the overlap of the corresponding spatial wave function is 1 when the different binding energies are applied to the focused molecule. Moreover, the $D$-wave components with small contribution do not significantly affect the magnetic moments of the $\Omega_{c}{D}_s^*$ molecule with $J^P={1}/{2}^{-}$, the $\Omega_{c}^*{D}_s^*$ molecule with $J^P={1}/{2}^{-}$, and the $\Omega_{c}^*{D}_s^*$ molecule with $J^P={3}/{2}^{-}$. More importantly, the magnetic moments of the $\Omega_{c}^*{D}_s^*$ molecular state with $J^P={1}/{2}^{-}$ and the $\Omega_{c}^*{D}_s^*$ molecule with $J^P={3}/{2}^{-}$ are obviously different, and the magnetic moment of the $\Omega_{c}^*{D}_s^*$ molecule with $J^P={3}/{2}^{-}$ is greater than that of the $\Omega_{c}^*{D}_s^*$ molecular state with $J^P={1}/{2}^{-}$. Therefore, the magnetic moment properties can serve as the vital
physical quantities to differentiate the $\Omega_{c}^*{D}_s^*$ molecular state with $J^P={1}/{2}^{-}$ and the $\Omega_{c}^*{D}_s^*$ molecule with $J^P={3}/{2}^{-}$.

Furthermore, several predicted $\Omega_{c}^{(*)}{D}_s^{(*)}$ molecules and conventional $\Omega_{cc}$ baryons may exist the identical quantum numbers and comparable masses. In the future, we will explore the electromagnetic characteristics of the conventional $\Omega_{cc}$ baryons, which can offer the vital information for distinguishing the $\Omega_{c}^{(*)}{D}_s^{(*)}$ molecules and the $\Omega_{cc}$ baryons.

\section{Summary}\label{sec4}

In recent decades, the investigation of the exotic hadronic states has become one of the most influential and important research issues in hadron physics. Thanks to the accumulation of the high-precision experimental data, more and more exotic hadronic states have been discovered in the high-energy physics experiments. In particular, LHCb has observed the $P_{\psi}^{N}$, $P_{\psi s}^{\Lambda}$, and $T_{cc}$ states, and their masses are very close to the thresholds of the corresponding two hadrons, which has inspired theorists to explain them in terms of the hidden-charm molecular pentaquarks and the doubly charmed molecular tetraquark. Obviously, these experimental observations provide a good opportunity for constructing the zoo of the doubly charmed molecular pentaquarks in the hadron spectroscopy. In Refs. \cite{Chen:2021kad,Yalikun:2023waw}, the authors have already predicted the existence of the doubly charmed molecular pentaquark candidates with the $\Sigma_c^{(*)} D^{(*)}$ and $\Xi_c^{(\prime,*)} D_s^{(*)}$ configurations.

To further complete the doubly charmed molecular pentaquark family, in this work we focus on the doubly charmed molecular pentaquark candidates of the $\Omega_{c}^{(*)}{D}_s^{(*)}$ configuration, which are new type of doubly charmed molecular pentaquarks containing most strange quarks. As the crucial information to search for them experimentally and the important inputs for the study of their properties, we first predict the mass spectra of the $\Omega_{c}^{(*)}{D}_s^{(*)}$-type doubly charmed molecular pentaquark candidates. In our concrete calculations, we utilize the OBE model to derive the effective potentials in the coordinate space of the $\Omega_{c}^{(*)}{D}_s^{(*)}$ systems, and then we try to find the loosely bound state solutions of the $\Omega_{c}^{(*)}{D}_s^{(*)}$ systems by solving the coupled channel Schr\"{o}dinger equation, where our analysis considers the contribution of the $S$-$D$ wave mixing effect and the coupled channel effect. Our quantitative analysis indicates that the $\Omega_{c}{D}_s^*$ state with $J^P={1}/{2}^{-}$, the $\Omega_{c}^*{D}_s^*$ state with $J^P={1}/{2}^{-}$, and the $\Omega_{c}^*{D}_s^*$ state with $J^P={3}/{2}^{-}$ can be considered as the most promising doubly charmed molecular pentaquark candidates, and the $\Omega_{c}{D}_s$ state with $J^P={1}/{2}^{-}$, the $\Omega_{c}^*{D}_s$ state with $J^P={3}/{2}^{-}$, and the $\Omega_{c}{D}_s^*$ state with $J^P={3}/{2}^{-}$ are the possible doubly charmed molecular pentaquark candidates, which are basically consistent with the theoretical prediction in Ref. \cite{Dong:2021bvy}. Furthermore, the corresponding spatial wave functions of the $\Omega_{c}^{(*)}{D}_s^{(*)}$ molecules can be obtained, which play a pivotal role in the investigation of their properties.

In order to provide more abundant suggestions for the construction of the mass spectra of the $\Omega_{c}^{(*)}{D}_s^{(*)}$-type doubly charmed molecular pentaquarks in the future experiments, we further explore  the radiative decays and the magnetic moments of the most promising doubly charmed molecular pentaquark candidates based on their mass spectra and spatial wave functions. In the realistic calculations, the constituent quark model is used by considering both the $S$-$D$ wave mixing effect and the coupled channel effect. Our obtained results indicate that (i) the radiative decay behaviors and the magnetic moments play an important role in reflecting the inner structures of the $\Omega_{c}^{(*)}{D}_s^{(*)}$-type doubly charmed molecular pentaquarks, (ii) the radiative decay behaviors of the $\Omega_c^* D_s^*|{1}/{2}^-\rangle \to \Omega_c D_s^*|{1}/{2}^-\rangle\gamma$ and $\Omega_c^* D_s^*|{3}/{2}^-\rangle \to \Omega_c D_s^*|{1}/{2}^-\rangle\gamma$ processes can provide the valuable suggestions for identifying the spin-parity quantum numbers of the $\Omega_{c}^*{D}_s^*$ molecular states, and (iii) the spin-parity quantum numbers of the $\Omega_{c}^*{D}_s^*$ molecular states can be distinguished by studying their magnetic moment properties.

Since the discovery of the charmonium-like state $X(3872)$, the past two decades have been a crucial period in the study of the exotic hadronic states. With the accumulation of higher statistical data at LHC \cite{Bediaga:2018lhg}, we have enough reason to believe that various types of doubly charmed molecular pentaquark candidates will be found at LHCb in the future. Thus, the mass spectra, the properties, and the production mechanisms of different types of doubly charmed molecular pentaquark candidates should receive more attentions from theorists at the present stage, which will expand our understanding of the doubly charmed molecular pentaquark candidates and provide the essential guidance to search for the doubly charmed molecular pentaquark candidates in the future experiments.

\section*{ACKNOWLEDGMENTS}

This work is supported by the China National Funds for Distinguished Young Scientists under Grant No. 11825503, the National Key Research and Development Program of China under Contract No. 2020YFA0406400, the 111 Project under Grant No. B20063, the National Natural Science Foundation of China under Grant Nos. 12247101, 12247155 and 12335001, the fundamental Research Funds for the Central Universities, and the project for top-notch innovative talents of Gansu province. F.L.W. is also supported by the China Postdoctoral Science Foundation under Grant No. 2022M721440.

\end{document}